\documentclass[useAMS,usenatbib]{mn2e}
\usepackage[pdfpagemode={UseOutlines},bookmarks,bookmarksopen,colorlinks,citecolor={blue},urlcolor={red}]{hyperref}
\usepackage[flushleft]{threeparttable}
\usepackage{booktabs,caption}
\usepackage{graphicx}
\usepackage{amsmath}
\usepackage{subfigure}
\usepackage{amssymb}
\usepackage{tabularx}
\usepackage{natbib}
\usepackage{float}
\usepackage{etoolbox}

\newcommand{\bea}{\begin{eqnarray}}
\newcommand{\eea}{\end{eqnarray}}

\makeatletter

\patchcmd{\NAT@citex}
  {\@citea\NAT@hyper@{%
     \NAT@nmfmt{\NAT@nm}%
     \hyper@natlinkbreak{\NAT@aysep\NAT@spacechar}{\@citeb\@extra@b@citeb}%
     \NAT@date}}
  {\@citea\NAT@nmfmt{\NAT@nm}%
   \NAT@aysep\NAT@spacechar\NAT@hyper@{\NAT@date}}{}{}

\patchcmd{\NAT@citex}
  {\@citea\NAT@hyper@{%
     \NAT@nmfmt{\NAT@nm}%
     \hyper@natlinkbreak{\NAT@spacechar\NAT@@open\if*#1*\else#1\NAT@spacechar\fi}%
       {\@citeb\@extra@b@citeb}%
     \NAT@date}}
  {\@citea\NAT@nmfmt{\NAT@nm}%
   \NAT@spacechar\NAT@@open\if*#1*\else#1\NAT@spacechar\fi\NAT@hyper@{\NAT@date}}
  {}{}

\makeatother

\title[Formation of MSSs in stellar clusters]{On the formation of compact, massive sub-systems in stellar clusters and its relation with intermediate mass black holes}

\author[M. Arca-Sedda]{M. ~Arca-Sedda$^{1}$\thanks{E-mail: m.arcasedda@gmail.com}\\
$^{1}$Dept. of Physics, University of Rome Sapienza, Piazzale Aldo Moro 5, I-00185, Rome (Italy)
}
\begin{document}
\date{Revised to 02-2015}

\pagerange{\pageref{firstpage}--\pageref{lastpage}} \pubyear{2015}

\maketitle

\label{firstpage}

\begin{abstract}
During their evolution, star clusters undergo mass segregation, by which the orbits of the most massive stars shrink, while the lighter stars move outwards from the cluster centre.
In this context, recent observations and dynamical modelling of several galactic and extra-galactic globular clusters (GCs) suggest that most of them show, close to their centre, an overabundance of mass whose nature is still matter of debate. 
For instance, many works show that orbitally segregated stars may collide with each other in a runaway fashion, leading to the formation of a very massive star or an intermediate mass black hole (IMBH) with a mass comparable to the observed mass excess. On the other hand, segregated stars can form a dense system if the IMBH formation fails.  
In this paper we study the early formation phase of a dense, massive sub-system (MSS) in several GCs models using a recently developed semi-analytical treatment of the mass segregation process.
In order to investigate how the MSS properties depend on the host cluster properties, we varied initial mass function (IMF), total mass, spatial distribution and metallicity of our models. Our results show how the IMF contributes to determine the final mass of the MSS, while the metallicity and the spatial distribution play a minor role. The method presented in this paper allowed us to provide scaling relations that connect the MSS mass and the host cluster mass in agreement with the observed correlation.
In order to follow the early formation stage of the MSSs and improve our statistical results, we performed several $N$-body simulations of stellar clusters with masses between $10^3$ and $2\times 10^5$ solar masses.
\end{abstract}


\section{Introduction}

Recently, a number of observations of globular clusters (GCs) suggested that their cores may contain much more mass then expected from earlier observations \citep{vandermarel10,noyola10}. 

As proposed by  several authors, these mass excesses can be ascribed to the presence of an intermediate mass black hole (IMBH) or a very massive star (VMS) close to the cluster centre, with masses in the range $10^2-10^4$ M$_\odot$ . 

For instance, observations and modelling of the innermost region of M15 seem to be consistent with the presence of a single object \citep{GEMA02,denBrock14}.
Furthermore, hints for IMBHs candidates have been found also in other GCs in the Milky Way \citep{NOGE08,Lutzgendorf13,Feldmeier13,Peuten14},
and in other galaxies (see for example \cite{mapelli08}), as in the case of G1 in M31 \citep{GERI02,GERI05,MillerJ12}, and the young massive cluster in the M82 irregular galaxy \citep{kaaret01,matsumoto01,usuda01}. 

Several mechanisms have been proposed for the birth and growth of IMBHs. For instance, \cite{MIHA02} pointed out that a stellar BH seed may grow slowly through occasional collisions and merging with other stars. 

Another possibility relies upon the mass segregation process, driven mainly by dynamical friction (df), which leads to an accumulation of mass within the cluster centre in form of orbitally decayed stars. 
In such a case, the most massive stars tend to concentrate toward the centre of the cluster, while the lighter component moves outward in an attempt to establish energy equipartition. As the mass segregation proceeds, the heaviest stars lose kinetic energy and reach the innermost region of the cluster, where they form a dense, contracting nucleus \citep{spitzer69,heggie03,bt}.

In this framework, some authors suggested that the shrinking nucleus collapses in a fraction of the relaxation time-scale \citep{zwart02,GURK}. The collapse facilitates a phase of runaway collisions among stars that leads to the formation of an IMBH \citep{Zwa99,zwart04,freitag06c,Goswami12,Lutz15,Giersz15}, whose gravitational field can significantly shape the dynamical evolution of the host cluster \citep{LutzBau13,Leigh14}.

Another possibility is that the observational evidence of a compact object in the centre of GCs can be interpreted as a dense, massive sub-system composed of dark remnants of heavy stars \citep{BAPZ03,vandermarel10,haggard13,lanzoni13,Kamann14}. 
In this case, the formation of binaries efficiently halts the contraction of the nucleus, preventing the formation of a VMS (or an IMBH). The evolution of the resulting sub-system of stars, which we refer to as MSS, will be likely dominated by two and three-body interactions \citep{BAPZ03,trani14}. 

A great effort toward the understanding of IMBHs formation and growth will likely come from the next generation of space-based gravitational waves (GWs) observatories, as the eLISA satellite \citep{amaro06,amaro09,mapelli10} but, currently, it is quite hard to discriminate between an IMBH and an MSS through observations.

In this paper, we investigate how the global properties of the host cluster affect the formation process of an MSS. Using a treatment for the dynamical friction process recently developed by \cite{ASCD14df}, we studied mass segregation in 168 models of GCs with different total masses, density distributions, initial mass functions (IMFs) and metallicities.

Our results suggest that the kind of stars which populate the MSS depends on the IMF and metallicity of the host cluster, while spatial distribution and stellar evolution affect significantly the MSS mass. Furthermore, we provide relations connecting the MSS mass with the total mass of the host cluster, showing that the best agreement with observations is achieved for power-law IMFs.

To follow the MSS formation process in detail, we performed direct $N$-body simulations of several GCs with masses in the range $10^3-10^5$ M$_\odot$. The time-scale needed to assembly an MSS, as well as the MSS mass and size, agrees with our semi-analytical results. 

In Section \ref{meth} the methodology used to derive MSS masses is introduced and discussed, in Section \ref{res} we investigate the properties of MSSs and provide scaling relations connecting the mass of MSSs and the host cluster, in Section \ref{nbody} we discuss the direct $N$-body simulations performed whereas in Section \ref{end} we draw the conclusions of this work.

\section{Numerical method}
\label{meth}

As database to compare with our theoretical results, we used observational data provided by \cite{LU13} (hereafter LU13), although some works seem to be at odds with them (see for example \cite{vandermarel10,haggard13,lanzoni13}). 
We will discuss the implications of different observational mass estimates on our results in the next section.
Table \ref{TABGC} summarises the main properties of the observed GCs.

\begin{table*}
\centering{}
\caption{Parameters of the observed GCs collected in LU13.}
\begin{center}
\begin{tabular}{cccccc}
\hline
\hline
ID & NAME & ${\rm Log} (M_{\rm GC}/$M$_\odot)$ & $\varepsilon_{M_{\rm tot}}$ & ${\rm Log} (M_{\rm BH}/$M$_\odot)$ & $\varepsilon_{M_{\rm BH}}$\\
\hline
G1      &           & $6.76$ & $ 0.02$& $4.25$&    $ 0.11$ \\
NGC104  &  47Tuc    & $6.04$ & $ 0.02$&  $<3.17$&  $ - $\\ 
NGC1851 &           & $5.57$ & $ 0.04$&  $<3.3$&   $ - $ \\
NGC1904 &  M79      & $5.15$ & $ 0.03$&  $3.47$&   $ 0.12$ \\
NGC2808 &           & $5.91$ & $ 0.04$&  $<4$&    $ - $ \\ 
NGC5139 &$\omega$Cen& $6.40$ & $ 0.05$&  $4.6$&    $ 0.08$ \\
NGC5286 &           & $5.45$ & $ 0.02$&  $3.17$ &  $ 0.24$  \\
NGC5694 &           & $5.41$ & $ 0.05$&  $<3.9$ &  $ -$  \\
NGC5824 &           & $5.65$ & $ 0.03$&  $<3.78$&   $ -$ \\
NGC6093 &   M80     & $5.53$ & $ 0.03$&  $<2.9$&   $ -$ \\
NGC6266 &   M62     & $5.97$ & $ 0.01$&  $3.3$ &  $ 0.18$ \\
NGC6388 &           & $6.04$ & $ 0.08$&  $4.23$ &  $ 0.18$ \\
NGC6715 &   M54     & $6.28$ & $ 0.05$&  $3.97$ &  $ 0.18$ \\
NGC7078 &   M15     & $5.79$ & $ 0.02$&  $<3.64$ &  $ -$ \\ 
\hline
\label{TABGC}
\end{tabular}
\end{center}
\begin{tablenotes}
\item Column 1: name of the cluster. Column 2: alternative name of the cluster. Column 3: mass of the cluster. Column 4: logarithmic error on the cluster mass. Column 4: mass of the central IMBH candidate. Column 5: logarithmic error on the IMBH candidate mass.
\end{tablenotes}
\end{table*}

To reach our aims, we need two important ingredients: i) a detailed description of the df process, which primarily drives heavy stars toward the host cluster centre, and ii) a reliable modelling of the host GC.

In order to estimate the final mass of MSSs, we sampled several GC models by varying their total mass, radius, initial mass function (IMF) and metallicity (Z). In particular, for each model we selected initial position, orbital eccentricity and mass of all its stars, which are crucial ingredients to determine how much stars contribute to the formation of an MSS.

\subsection{Dynamical friction}
\label{df}

A massive body traversing a sea of lighter particles suffers a dynamical braking that drags it toward the centre of the host system \citep{CVN43,Cha43I,Cha43b}. 

This mechanism, called dynamical friction (df), arises directly from two-body encounters and plays a significant role in shaping the evolution of astrophysical systems on very different scales \citep{Bekenstein,Trem76,Dolc93,Milos01,Gual08,Ant13,ASCD14nc,ASCD15he}. 

In a pioneriing paper, \cite{Cha43I} provided the timescale over which a body of mass $m_*$
that moves on an orbit with initial apocenter $r_*$ and initial velocity $v_*$, reaches the innermost region of its host system \citep{bt}:
\begin{equation}
\tau_{\rm df}(\mathrm{Myr})=\frac{1.9\times 10^4}{{\rm log} \Lambda}\left(\frac{r_*}{5~{\rm kpc}}\right)^2\left(\frac{v_*}{200~{\rm kms^{-1}}}\right)\left(\frac{10^8~{\rm M}_\odot}{m_*}\right),
\label{tdf}
\end{equation}
where, for star clusters, we assume ${\rm log} \Lambda\sim 10$ for the usual Coulomb logarithm.

Many works attempted to generalise the Chandrasekhar's work to axisymmetric and triaxial systems \citep{Bin77,OBS,Pes92}, and to systems characterised by cusped density profiles \citep{Me06,Vicari07,AntMer12,ASCD14df}.

In particular, \cite{ASCD14df} developed a reliable treatment of df particularly well suited to describe the motion of massive bodies in both cusped and cored density profiles. 
Furthermore, they provided an useful formula for the df time-scale, recently updated in \cite{ASCD15he}:
\begin{equation}
\tau_{\rm df} (\mathrm{Myr}) = T_{\rm ref}\sqrt{\frac{r_{\rm GC}^3}{M_{\rm GC}}}
g(e,\gamma)\left(\frac{m_*}{M_{\rm GC}}\right)^{-0.67}\left(\frac{r_*}{r_{\rm GC}}\right)^{1.76},
\label{eq2}
\end{equation}

where $T_{\rm ref} = 0.3\sqrt{10^{11}{\rm M}_\odot / 1{\rm kpc}^3}$ and $e$ identifies the initial eccentricity of the orbit
\begin{equation}
e = \frac{r_*-r_p}{r_*},
\end{equation} 
with $r_p$ the pericentral distance from the cluster centre.
The function $g(e)$, instead, links $e$ and the slope of the density profile, $\gamma$, through the relation
\begin{equation}
g(e,\gamma)=(2-\gamma)\left[a_1\left(\frac{1}{(2-\gamma)^{a_2}}+a_3\right)(1-e) + e\right],
\end{equation}
with $a_1= 2.63 \pm 0.17$, $a_2 = 2.26 \pm 0.08$ and $a_3=0.9 \pm 0.1$.

Considering a typical GC characterised by a cored density profile ($\gamma=0$), with a mass $M_{\rm GC}\sim 10^6$ M$_\odot$ and a scale radius $r_{\rm GC} = 1$ pc, it easy to find through Equation \ref{eq2}
\begin{equation}
\tau_{\rm df}\lesssim 0.6 ~{\rm Gyr},
\end{equation}
for a massive star ($m_*\sim 25$ M$_\odot$) that moves on a circular orbit at $r_*=5$ pc from the GC centre.

Hence, a population of heavy stars may sink to the centre of the host GC in few Gyr, leading to a significant accumulation of mass in its innermost region.

\subsection{Sampling method}
\label{smpl}

To provide reliable estimates of the amount of stars that have sunk to the cluster centre in a given time, we selected isolated cluster models with masses in the range $10^3-3\times 10^6$ M$_\odot$, which differ each one in spatial distribution, IMF and metallicity. 
Furthermore, we included the SSE package \citep{hurley} in our statistical code, in order to take in account stellar evolution, which causes stellar mass loss and may alter the df time.
The properties of our GC models are discussed in detail in the following sections.

\subsubsection*{Spatial distribution}

We sampled the positions of each star according either to an uniform spatial distribution, in which the density of stars does not depend on the spatial coordinates, or to a cored $\gamma$-profile ($\gamma=0$) \citep{deh}. 

Furthermore, we assigned to stars' orbits an orbital eccentricity, $e$, in the range $e=0$ (circular orbits) and $e=1$ (pure radial orbits) according to a flat distribution.
We verified that the use of different sampling methods for $e$ does not affect significantly the global results, unless all the stars move on circular, or purely radial, orbits.

Although an uniform sphere evolves over few dynamical times and the treatment presented here do not account for the time evolution of its global structure, we will show in the next section that our semi-analytical estimates well reproduce the accumulation of mass in cluster models characterised by a constant density by comparing them with direct $N$-body simulations.

\subsubsection*{Initial mass function and metallicity of the cluster}

We assigned masses to the stars in the range $0.1$ and $100$ M$_\odot$ according to either a flat IMF, a Salpeter IMF \citep{salp} or a Kroupa IMF \citep{krp}. 

Moreover, to highlight the effects of different metallicities ($Z$) on the formation of an MSS, we assigned to stars either a solar metallicity $Z=0.02$, or the typical metallicity of old globular clusters, $Z=0.0004$.
\\

At the end, we gathered a total sample of 168 models, whose main parameters are summarised in Table \ref{MODEL}. 

Moreover, we made 100 realisations of each cluster, in order to filter out statistic fluctuations, providing an averaged value of the MSS mass along with an estimate of the corresponding standard deviation.

We grouped clusters having different masses but the same global properties, labelling them with the letter A or B and a number between 1 and 6. 
 
It should be noted that such a choice brings together systems in which the typical time-scale over which an MSS form can be very different, since it is directly connected to the total mass of the host cluster.

In particular, our grouping choice gathers clusters with ages greater than the typical time-scale over which an MSS forms.

Letter A refers to models with solar metallicity, whereas letter B indicates metal-poor models.
The number, instead, depends on the kind of IMF and the spatial distribution of the stars. Each number represents a combination of the IMF and the spatial distribution that characterises the model. For instance, number 1 refers to models with a Kroupa IMF and an uniform spatial distribution.
 
As pointed out above, we selected three IMFs: flat (denoted with letter F in Table \ref{MODEL}), Kroupa (K) and Salpeter (S). The spatial distributions considered, instead, are uniform (U) or cored $\gamma$-distribution (D).

\begin{table}
\caption{}
\centering{Parameters of the cluster models.}
\begin{center}
\begin{tabular}{ccccc}
\hline
\hline
Model & IMF & $\rho(r)$ & $Z$ & $M_{\rm GC} ~($M$_\odot)$\\
\hline  
A1	& K & U & $ 0.02 $ & $ 10^3-3\times10^6 $\\      
A2	& K & D & $ 0.02 $ & $ 10^3-3\times10^6 $\\     
A3	& S & U & $ 0.02 $ & $ 10^3-3\times10^6 $\\       
A4	& S & D & $ 0.02 $ & $ 10^3-3\times10^6 $\\       
A5	& F & U & $ 0.02 $ & $ 10^3-3\times10^6 $\\        
A6	& F & D & $ 0.02 $ & $ 10^3-3\times10^6 $\\        
B1  & K & U & $ 0.0004$ & $ 10^3-3\times10^6 $\\      
B2  & K & D & $ 0.0004$ & $ 10^3-3\times10^6 $\\        
B3  & S & U & $ 0.0004$ & $ 10^3-3\times10^6 $\\         
B4  & S & D & $ 0.0004$ & $ 10^3-3\times10^6 $\\         
B5  & F & U & $ 0.0004$ & $ 10^3-3\times10^6 $\\         
B6  & F & D & $ 0.0004$ & $ 10^3-3\times10^6 $\\        
\hline
\label{MODEL}
\end{tabular}
\end{center}
\begin{tablenotes}
\item Note. Column 1: name of the model. Column 2: IMF used to sample masses of the stars: Kroupa (K), Salpeter (S) and flat (F). Column 3: spatial density profile of the stars: uniform (U) and Dehnen (D). Column 4: metallicity of the cluster. Column 5: masses of the cluster models.
\end{tablenotes}
\end{table}

\subsection{Evaluation of MSS masses and sizes}
\label{MSSeva}

A star undergoes orbital decay only if its mass exceeds the mean mass of the background stars $\left\langle  m_*\right\rangle $ \citep{Cha43I}. In particular, several works pointed out that dynamical friction affects significantly the motion of bodies with masses $m_*\geq K\langle m_* \rangle$, where $K$ is a parameter in the range $1-50$, depending on the method used in describing df \citep{colpi99,ASCD14df,alessandrini14,miocchi15}. In this paper, we assumed $K=30$, a value well supported by previous theoretical and numerical results \citep{ASCD14df}.

Furthermore, mass loss process can alter the decay process, leading to a significant increase of the df time, at least for the most massive stars.

In particular, it is worth nothing that stars with masses above $30$ M$_\odot$ lose most of their initial mass in $< 10$ Myr, leading to a significant decrease of the df efficiency well before they reach the cluster centre. Indeed, an inversion of Equation \ref{eq2} allows to guess the initial position that a star should have to reach the cluster centre within a time $t$:

\begin{equation}
r(t,m_*)= \left(\frac{t}{\mathcal{T}}\right)^{0.57}\left(\frac{m_*}{M_{\rm GC}}\right)^{0.38}r_{\rm GC},
\label{tdfrad}
\end{equation}
where $\mathcal{T}=T_{\rm ref}g(e,\gamma)\left(r_{\rm GC}^3/M_{\rm GC}\right)^{1/2}$.

For example, a star with $m_*=30$ M$_\odot$ that moves in a typical GC with $M_{\rm GC}= 10^6$ M$_\odot$, $r_{\rm GC} = 1$ pc and $\gamma = 0$, should have an initial position $r(t,m_*)\simeq 0.6$ pc to reach the GC centre within $10$ Myr.

Hence, we took in account how the df time-scale changes as a consequence of the stellar evolution proceeding in the following way. We divided the time $t$ in sub-intervals $\delta t_i$ equally spaced such that $\sum_i \delta t_i = t$, keeping each interval small enough to follow in detail the evolution of the star mass as a function of the time, i.e. $\delta t_i\lesssim 1$ Myr. For each value of $\delta t_i$, we evaluated the updated mass of the star through the SSE package, $m_{*i}$, and the df time that corresponds to such a mass, $\tau_{df,i}$, using Equation \ref{eq2}.

We defined the final df time-scale of a star as the weighted-average value of all the df times $\delta t_i$ over the total time interval: 
\begin{equation}
t_{\rm df} = \frac{\sum_i m_{*i}\tau_{df,i}}{\sum_i m_{*i}},
\label{eq2cor}
\end{equation}
using its actual mass $m_{*i}$ as weight.

At the end, the total mass of the MSS is then evaluated as the sum of the masses of those stars with $t_{\rm df}\leq t$ {and $m_*\geq K\langle m \rangle$, plus the masses of stars with initial position smaller than the expected MSS size, $r_*<r_{\rm MSS}$}.

Furthermore, this method allows to provide hints on the size of the MSS. Indeed, as pointed out by several authors \citep{kal72, readcole, Gual08, AntMer12,ASCD14df}, df stops when the star, with mass $m_*$, approaches a distance from the cluster centre that encloses an amount of mass roughly equal to $m_*$. This critical radius is commonly called ``stalling radius'', $r_{\rm MSS}$.
Within $r_{\rm MSS}$, the motion of the star is mainly dominated by random encounters, while the df action becomes negligible.
Since the greater the star mass the greater the stalling radius, we can define as MSS size the $r_{\rm MSS}$ of the most massive star ($m_* = 100$ M$_\odot$).

For a $\gamma$-profile, whose radial mass distribution is given by:
\begin{equation}
M_{\rm GC}(r) = M_{\rm GC}\left(\frac{r}{r+r_{\rm GC}}\right)^{3-\gamma},
\label{MrDEH}
\end{equation}
this ``stalling radius'' can be obtained by inverting the latter equation, leading to:
\begin{equation}
r_{\rm MSS} = r_{\rm GC}\frac{\mathcal{M}}{1-\mathcal{M}},
\label{r_st}
\end{equation}
with $\mathcal{M} = (m_*/M_{\rm GC})^{1/(3-\gamma)}$. 

For an uniform distribution, instead, it is easy to find:
\begin{equation}
r_{\rm MSS} = R_{\rm GC}\left(\frac{3m_*}{4\pi M_{\rm GC}}\right)^{1/3},
\label{r_st2}
\end{equation}
being $R_{\rm GC}$ the total radius of the cluster. 

\begin{figure}
\includegraphics[width=8cm]{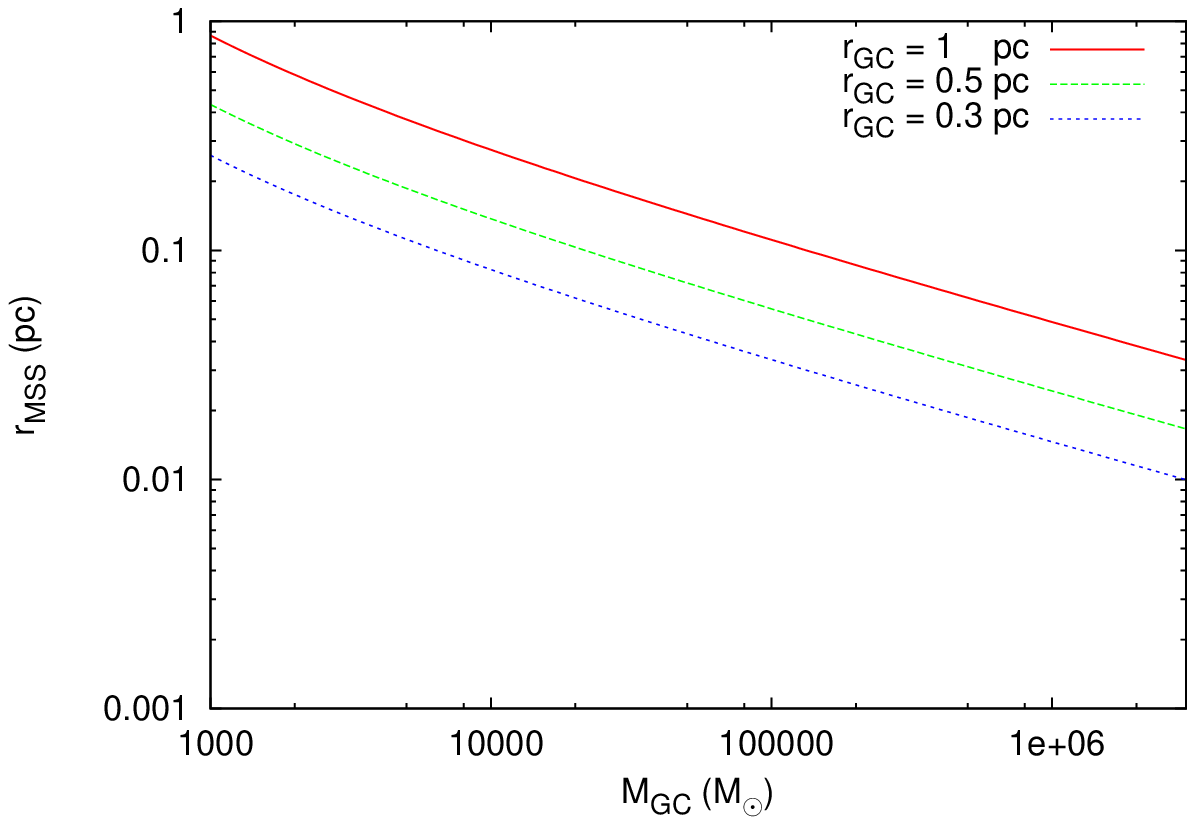}\\
\includegraphics[width=8cm]{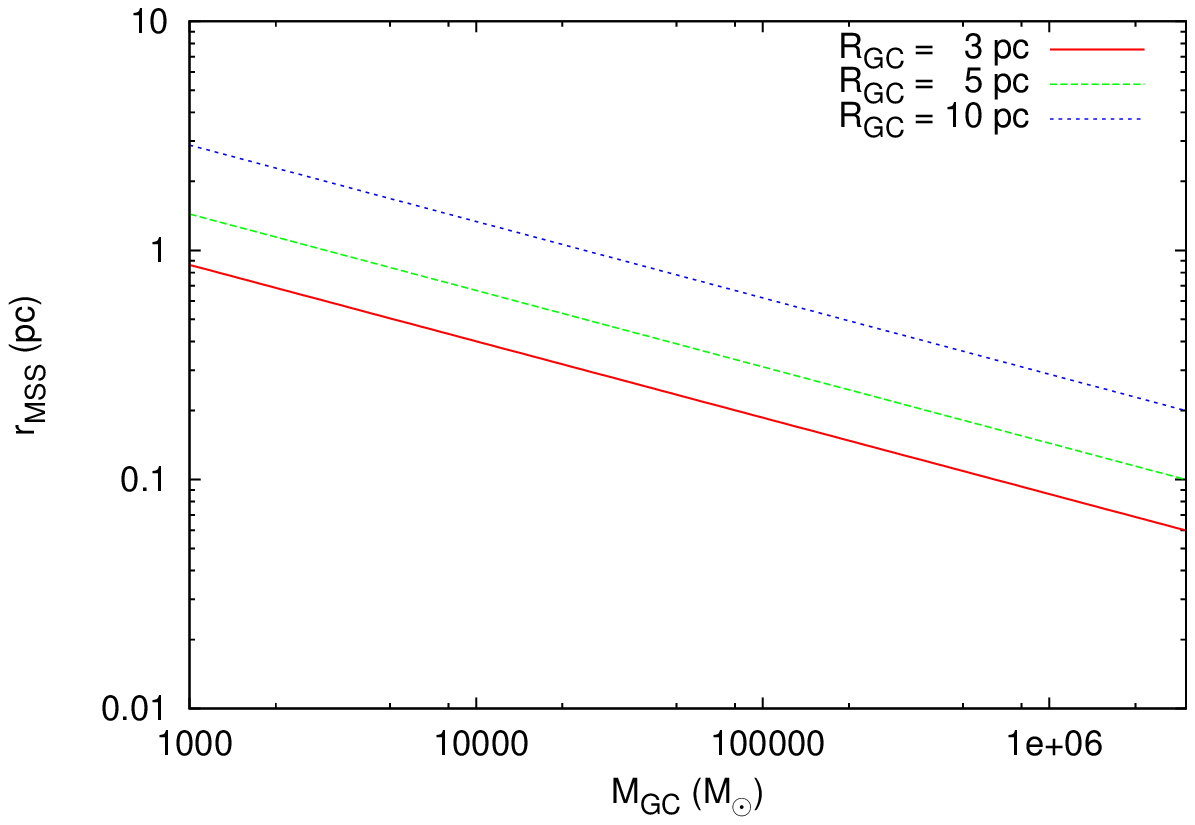}
\caption{Size of the MSSs obtained using Equation \ref{r_st} for models with a $\gamma$ density profile (top panel) and Equation \ref{r_st2} for models with an uniform density profile (bottom panel).}
\label{rstal}
\end{figure}

Figure \ref{rstal} shows the MSS sizes as a function of the GC mass for different values of the cluster scale radius, $r_{\rm GC}$, in the case of a $\gamma$ density profile and in the case of an uniform density profile.

It is quite evident that the heavier the host cluster the smaller the MSS radius, which can reach values below $0.1$ pc for $M_{\rm GC}>10^6$ M$_\odot$. It is worth noting that most of the observed mass excesses are enclosed within a region whose extension is $<0.5$ pc, quite close to the radii of our MSSs \citep{vandermarel10,haggard13,lanzoni13}.

The time-scale over which the formation of an MSS takes place depends on the total mass of the host cluster. In particular, 
as the stars' orbits shrink to the cluster centre, the MSS mass increases, reaching a saturation value over a time-scale that represents an upper limit to the relaxation time of the system \citep{bt,HENON,freitag06b}. Hence, it is evident that our choice to group models with the same properties but different masses implies that the time-interval over which their MSS should have formed can be very different.

Though our methodology is substantially blind to their subsequent evolution, we examine in the following two possible fates for the MSSs for the sake of completeness:
\begin{itemize}
\item heavy stars accumulate into the cluster centre in a time much shorter than the stellar mass loss time-scale. In this case, the contraction of the MSS drives the inner region of the cluster toward core collapse, thus facilitating a runaway collision phase \citep{Zwa99,zwart02,zwart04,freitag06c,Lutz15}, even boosted by binaries formed during the collapse process \citep{Zwa99,zwart02,freitag06d,gaburov08} and by primordial binaries \citep{zwart07}, which leads to the formation of an IMBH \citep{Zwa99,zwart04,GURK}. In this scenario, the MSS can represent the total reservoir of mass available in form of stars to build-up the IMBH through runaway collisions;
\item stellar mass loss occurs over a time comparable to the MSS formation. In this case, the formation of binaries and multiple systems within the MSS halts efficiently the contraction process, thus quenching stellar collisions \citep{angeletti77,chernoff90,zwart07,vesperini09,lamers13,mapelli13,fujii14}. In this case, the inner region of the cluster undergoes a series of contraction and re-expansions called gravothermal oscillations \citep{bettwieser84,cohn89,makino96} that drives an expansion of the MSS by a factor up to $\sim 2.5$ \citep{trani14}. 
It is worth noting that our estimates of the MSSs mass represent quite well the expected mass of these small cores.
\end{itemize}

\section{Results}
\label{res}

Using the approach described above, we examine in this section the stellar content of the newly born MSSs, providing also correlations that link the masses of MSSs and those of their host clusters.

\subsection{Properties of MSSs}
\label{proMSS}

In the following, we use models A2 and B2 as reference cases, in order to highlight the differences arising from the choice of different metallicities, and models A1 and A2 to highlight, instead, the effects connected to different spatial distributions of the host cluster.
As pointed out in the previous sections, we would demonstrate here that the mass excess observed in the centre of several GCs is likely related to a sub-system of orbitally segregated stars, not necessarily to an IMBH.

\begin{figure}
\centering
\includegraphics[width=8cm]{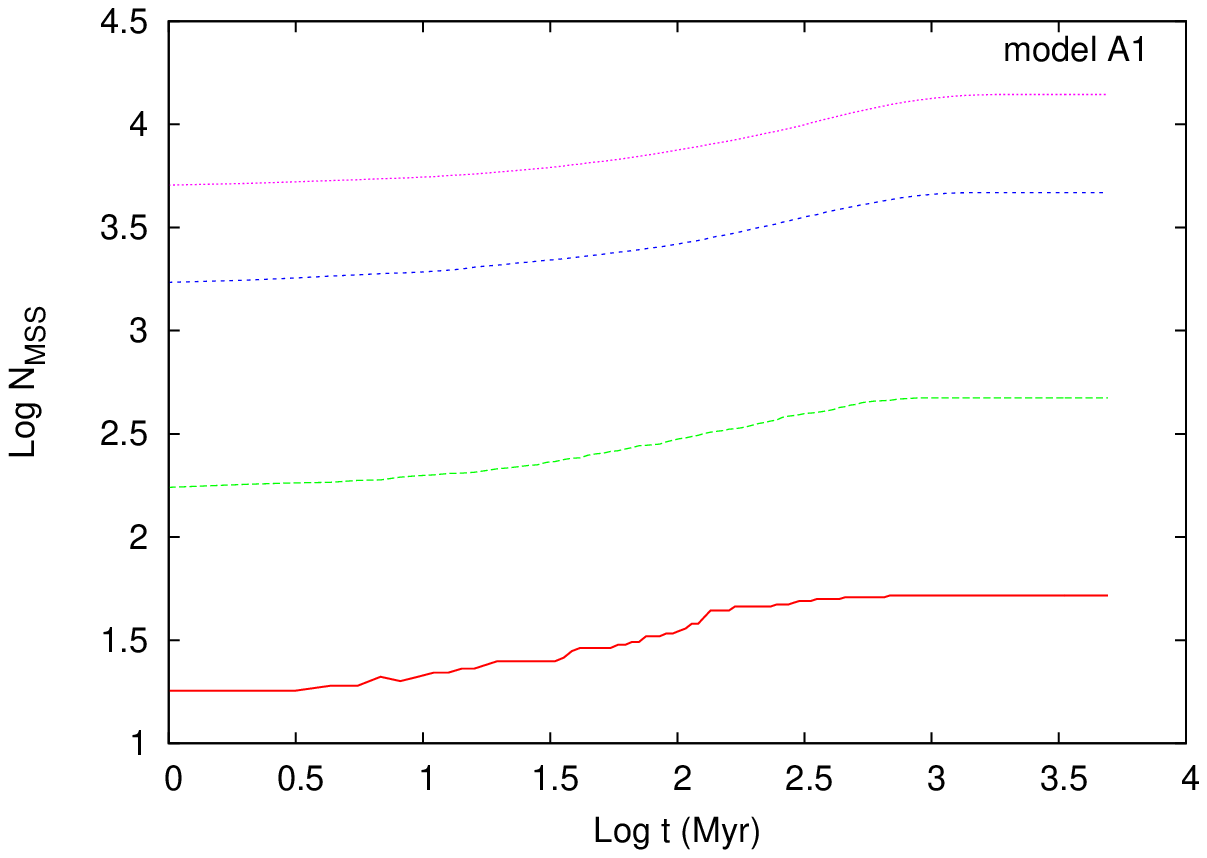}\\
\includegraphics[width=8cm]{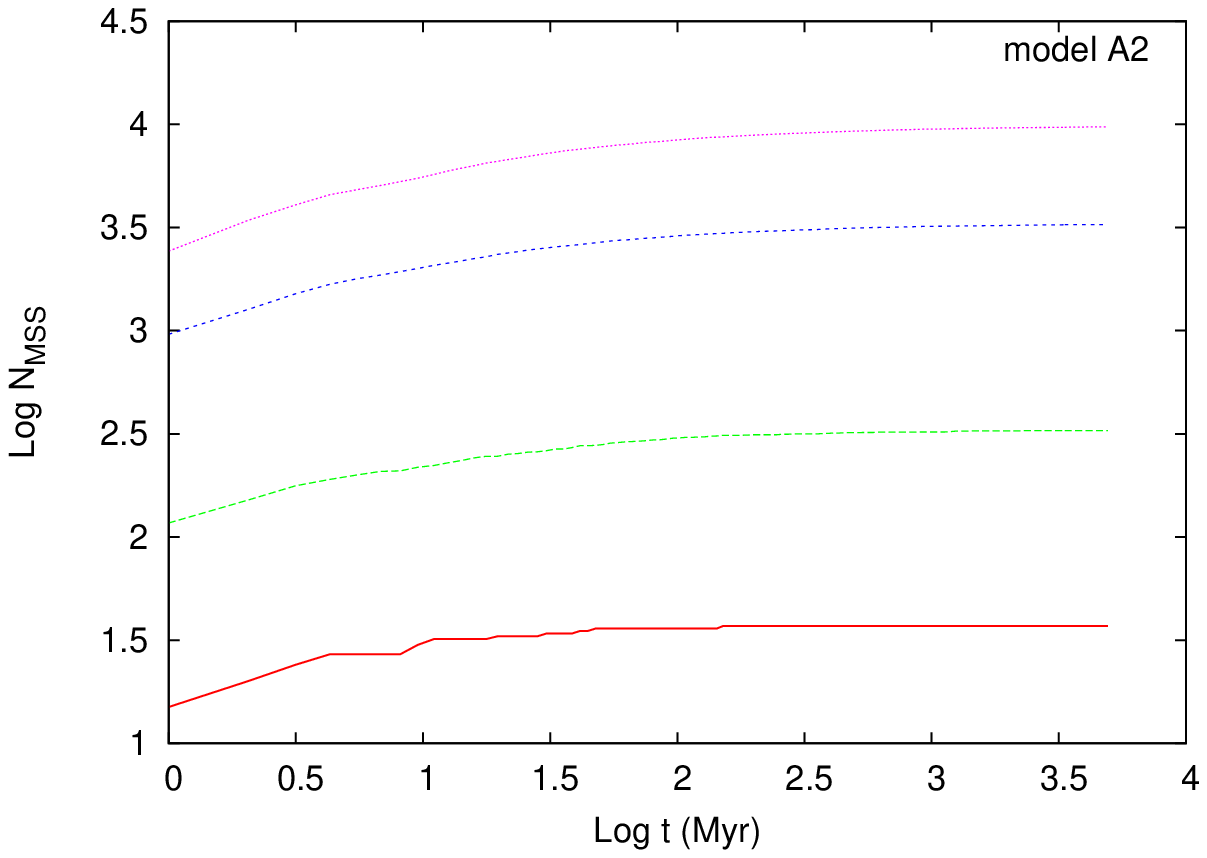}\\
\includegraphics[width=8cm]{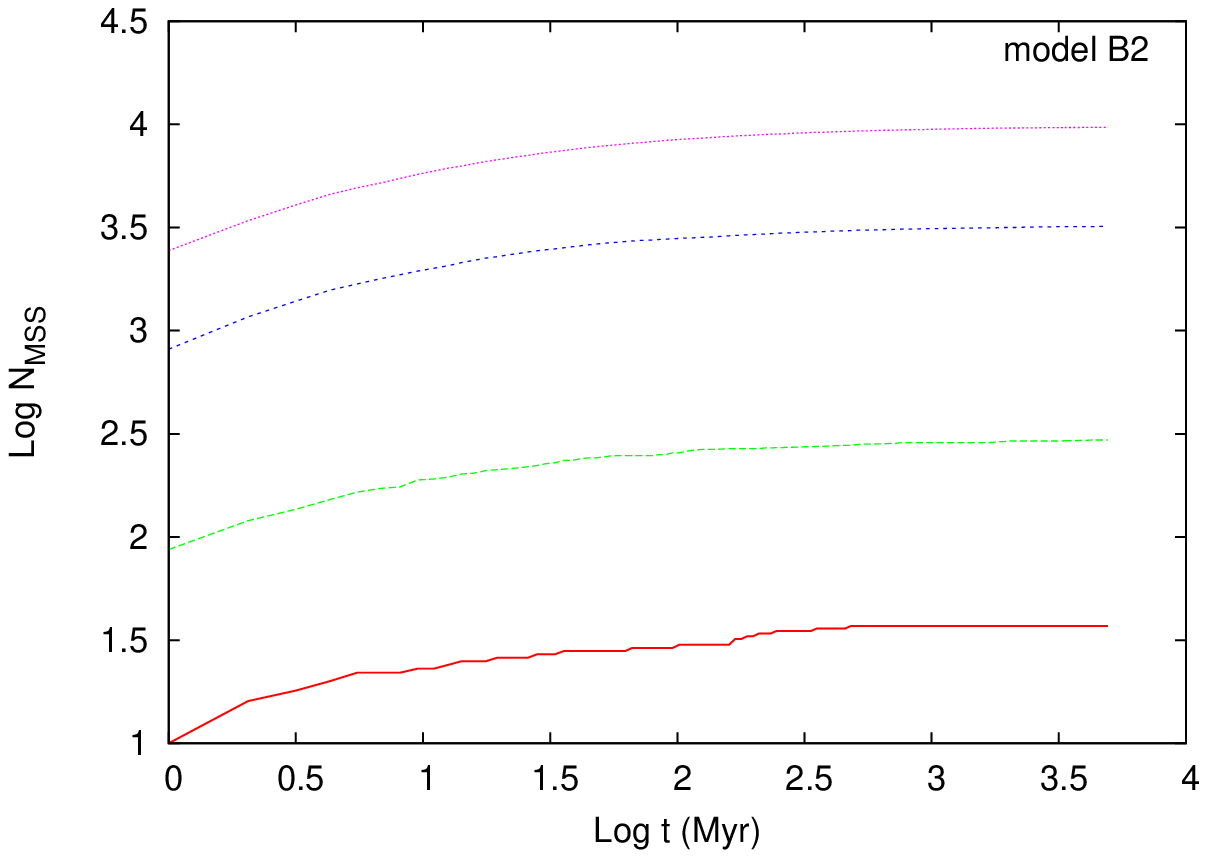}
\caption{Number of orbitally segregated stars as a function of the time for different masses of the host GC. Each panel refers to a different model, as indicated. Within each panel, instead, from bottom to top, lines refer to a cluster model with a mass $M_{\rm GC}=10^4-10^5-10^6-3\times 10^6$ M$_\odot$, respectively.}
\label{Fn1}
\end{figure}

Figure \ref{Fn1} shows the number of orbitally segregated stars, $N_{MSS}$, as a function of the time for models A1, A2 and B2 and for different GC total masses.

The growth process is characterised by two distinct phases: one more rapid, lasting up to $0.1-0.5$ Gyr, during which stars with nearly radial orbits or with small apocentres segregate fastly to the centre, and a second, slower phase, during which the main contribution to the MSS growth is given by stars that move on a more peripheral region of the cluster, thus reaching occasionally the GC centre.

\begin{figure}
\centering
\includegraphics[width=8cm]{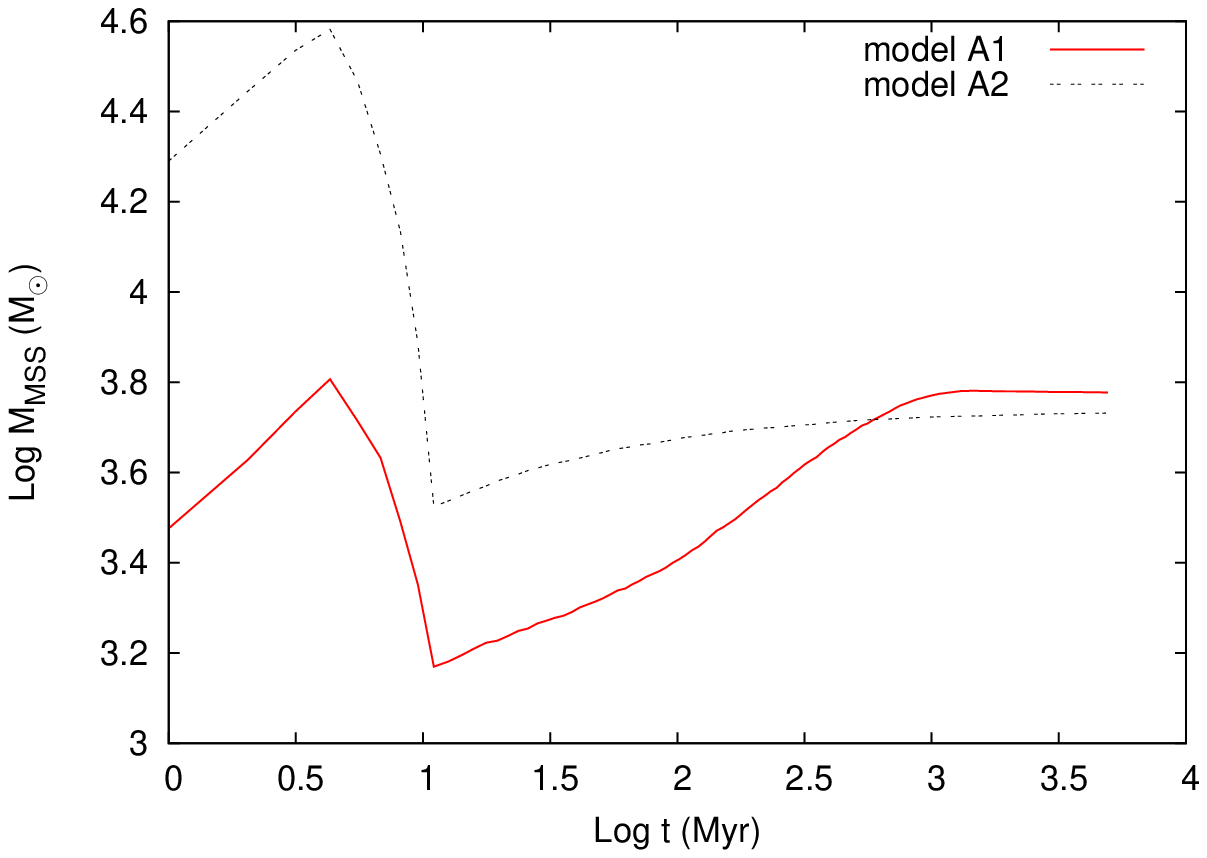}\\
\includegraphics[width=8cm]{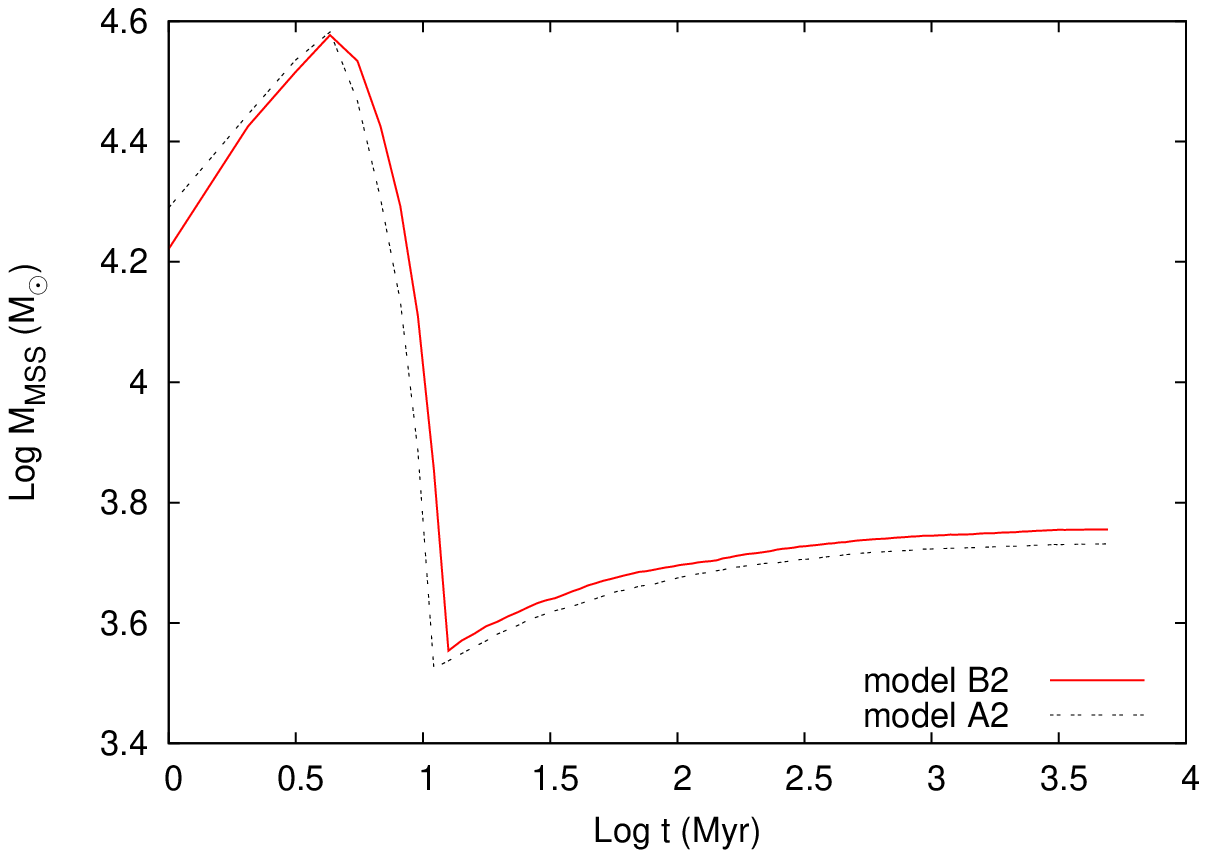}\\
\caption{Comparison between the mass deposited in the centre of a cluster with $M = 10^6$ M$_\odot$ for models A1 and A2 (top panel) and for models B2 and A2 (bottom panel), respectively.}
\label{Fn2}
\end{figure}

However, since stars lose mass during their evolution, the parameter $N_{MSS}$ cannot be used to provide informations about the MSS properties. 
In order to investigate how the MSS mass increases in time, we show in Figure \ref{Fn2} the total amount of mass accumulated within the cluster centre for the same models, considering a GC model with mass $M = 10^6$ M$_\odot$.

Looking at the figure, three different phases are clear: the deposited mass increases until it reaches a maximum value, then the mass accumulation process undergoes a rapid decrease phase which last $2-3$ Myr and finally it rises smoothly toward a saturation value.
The first phase corresponds to the decay of the most massive stars located in an inner region of the GC. Then, the mass decrease during the second phase as a consequence of the mass lost by segregated stars. Finally, the last stage is mostly determined by the deposit of stars with initial orbits located in the outer shells of the GC.

Comparing the MSS mass growth for model A1 and A2 in the case of $M_{\rm GC}=10^6$ M$_\odot$, it is evident that the deposited mass in model A2 is initially greater than in model A1. This is due to the fact that the A2 density profile is more concentrated,
thus implying that in this case stars move on orbits with smaller apocentres (on average) and therefore smaller df times.
On the other hand, over a time $\gtrsim 2$ Gyr, the deposited mass reach comparable values both in model A1 and A2. 

This is mainly due to the fact that the density profile of model A2 scales as $\rho(r)\propto (r+r_{\rm GC})^{-4}$ and, therefore, stars with initial apocenter $r\gg r_{\rm GC}$ travel in a low dense environment in which df is highly suppressed. On the other hand, since model A1 has an uniform density profile, stars moving on farther orbits may reach the GC centre since df acts efficiently also on larger length-scales.

Comparing models A2 and B2, instead, we found that the MSS mass evolution is very similar. In particular, in model B2 the mass of the MSS reaches a saturation value greater than in  model A2. This is due to the fact that stars with masses above $10$ M$_\odot$ having low metallicities reach final masses up to $15\%$ greater than stars with solar metallicity.
Such a difference may be even greater depending on the kind of stellar evolution recipes considered, as pointed out by several authors \citep{brocato,mapelli13,ziosi,Spera15b}.

\begin{figure}
\centering
\includegraphics[width=8cm]{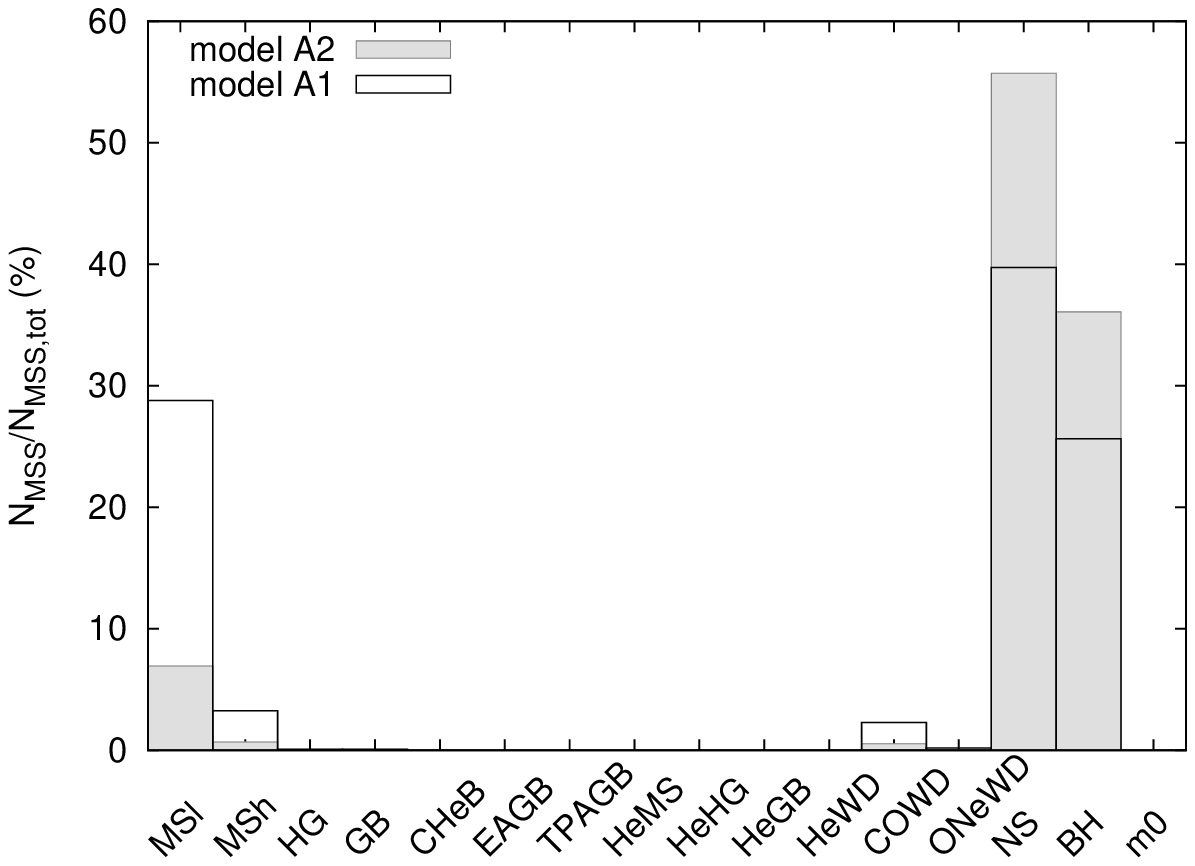}\\
\includegraphics[width=8cm]{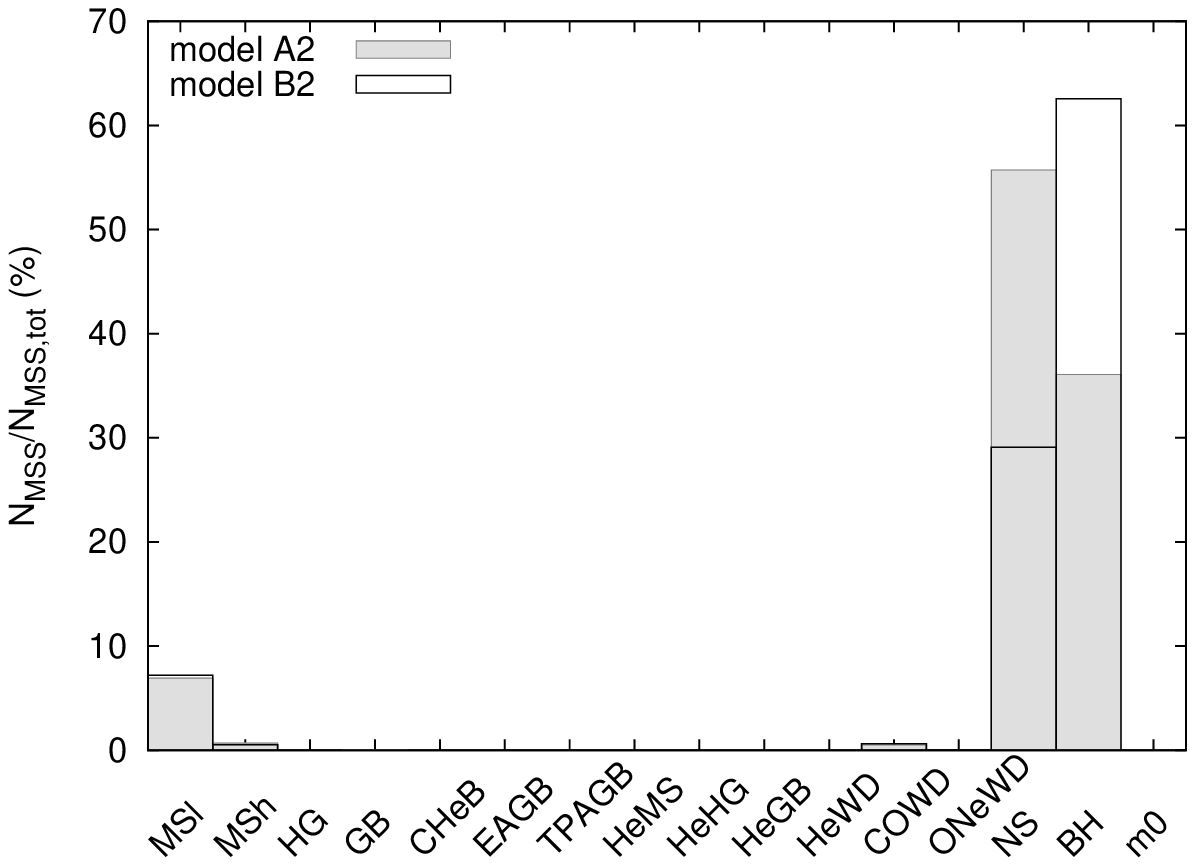}
\caption{Number of orbitally decayed stars, within $t=6$ Gyr, of a given stellar type over the total number of decayed stars in models A2 and A1 (top panel) and in models A2 and B2 (bottom panel) for a cluster with $M=10^6$ M$_\odot$.}
\label{Fn4}
\end{figure}

We can also provide hints about the stellar content of the MSSs, looking at stars form them.
In particular, Figure \ref{Fn4} shows the fractional number of stars that should form the MSS of a GC with a total mass $M_{\rm GC}=10^6$ M$_\odot$ and an age $t\simeq 6$ Gyr.
Stellar types are defined as in \cite{hurley}, and are listed in Table \ref{cod}.

The comparison between models with different metallicities (A2 and B2) highlights that a cluster with lower values of $Z$ will host an MSS likely dominated by BHs, which are more than $60\%$ of the total number of orbitally segregated stars, while neutron stars (NSs) seem to be the most common stars in a MSS of a cluster with solar metallicity. 
Furthermore, also the spatial distribution of stars seems to be important in determining the stellar composition of the MSS. Indeed, the fraction of BHs and NSs is smaller in models with an uniform spatial distribution (A1) with respect to the models with a $\gamma$ density profile (A2). Regarding model A1, we found that also a population of low main sequence stars (MSl) with inital apocentres quite close to the centre of the host GC, can contribute significantly to the total MSS mass.

\begin{table}
\caption{}
\centering{Evolution phases.}
\begin{center}
\begin{tabular}{ccc}
\hline
\hline
REF N.  &  NAME     &  Description \\
\hline
   0    &  MSl      &  MS stars with $M\leqslant0.7$M$_\odot$ \\
   1    &  MSh      &  MS stars with $M > 0.7$M$_\odot$\\
   2    &  HG       &  Hertsprung Gap\\
   3    &  GB       &  First GB\\
   4    &  CHeB     &  Core Helium Burning\\
   5    &  EAGB     &  Early AGB\\
   6    &  TPAGB    &  Thermally Pulsing AGB\\
   7    &  HeMS     &  Naked Helium star MS\\
   8    &  HeHG     &  Naked Helium star Hertsprung Gap\\
   9    &  HeGB     &  Naked Helium star GB\\
  10    &  HeWD     &  Helium White Dwarf\\
  11    &  COWD     &  Carbon/Oxygen White Dwarf\\
  12    &  ONeWD    &  Oxygen/Neon White Dwarf\\
  13    &  NS       &  Neutron Star\\       
  14    &  BH       &  Black Hole\\         
  15    &  m0       &  massless remnant\\         
\hline
\end{tabular}
\label{cod}
\end{center}
\begin{tablenotes}
\item Column 1: reference number used in \cite{hurley}. Column 2: abbreviation. Column 3: stellar evolution phase.
\end{tablenotes}
\end{table}

Equation \ref{eq2} can be used to provide some hints about the global time evolution of the stellar distribution of the stars in the cluster.
In particular, the position, $r$, of a star at a given time $t$ can be obtained by solving the equation: 
\begin{equation}
\tau_{\rm df}(r_*)-\tau_{\rm df}(r) = t,
\end{equation}
whose explicit solution is given by:
\begin{equation}
r(t) = r_* \left[1-\frac{t}{\tau_{\rm df}(r_*)}\right]^{0.57}.
\label{trajectory}
\end{equation} 

Hence, Equation \ref{trajectory} can be used to investigate how the spatial distribution of stars evolves in time.

\begin{figure}
\includegraphics[width=8cm]{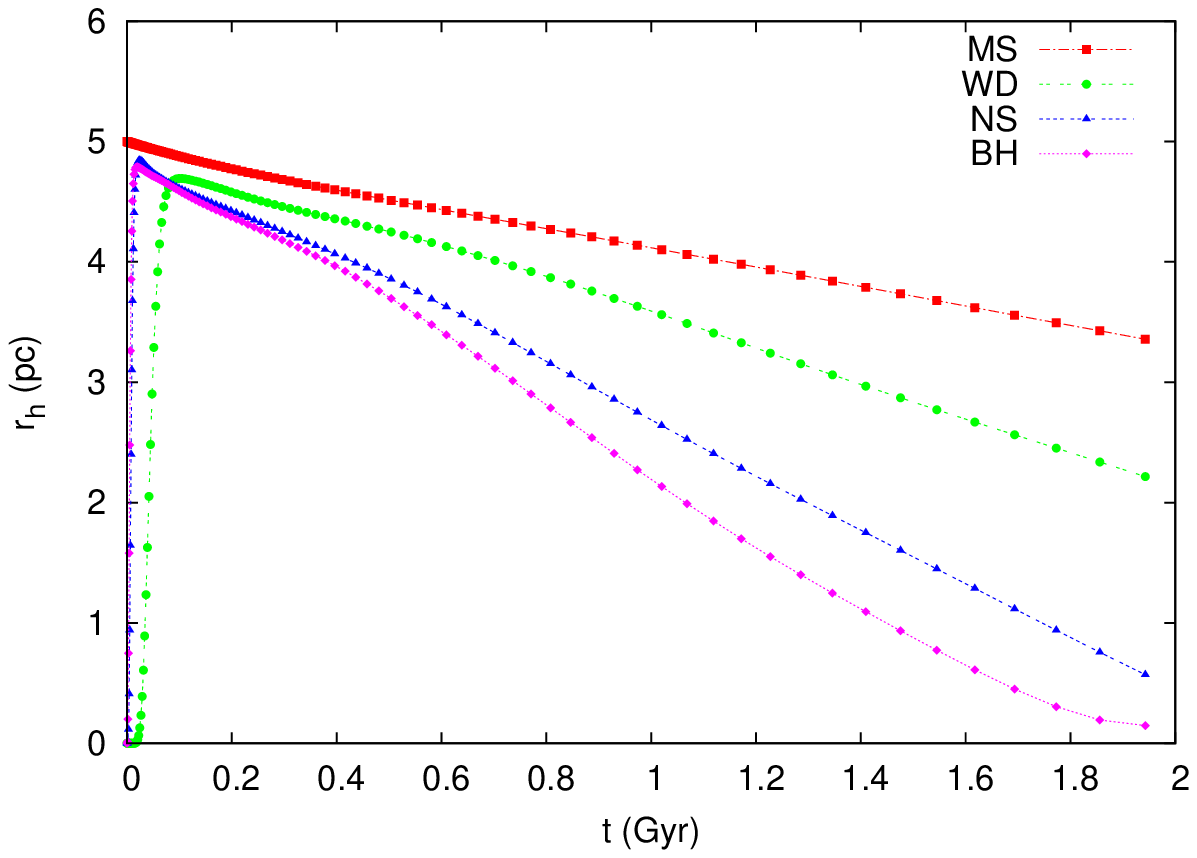}\\
\includegraphics[width=8cm]{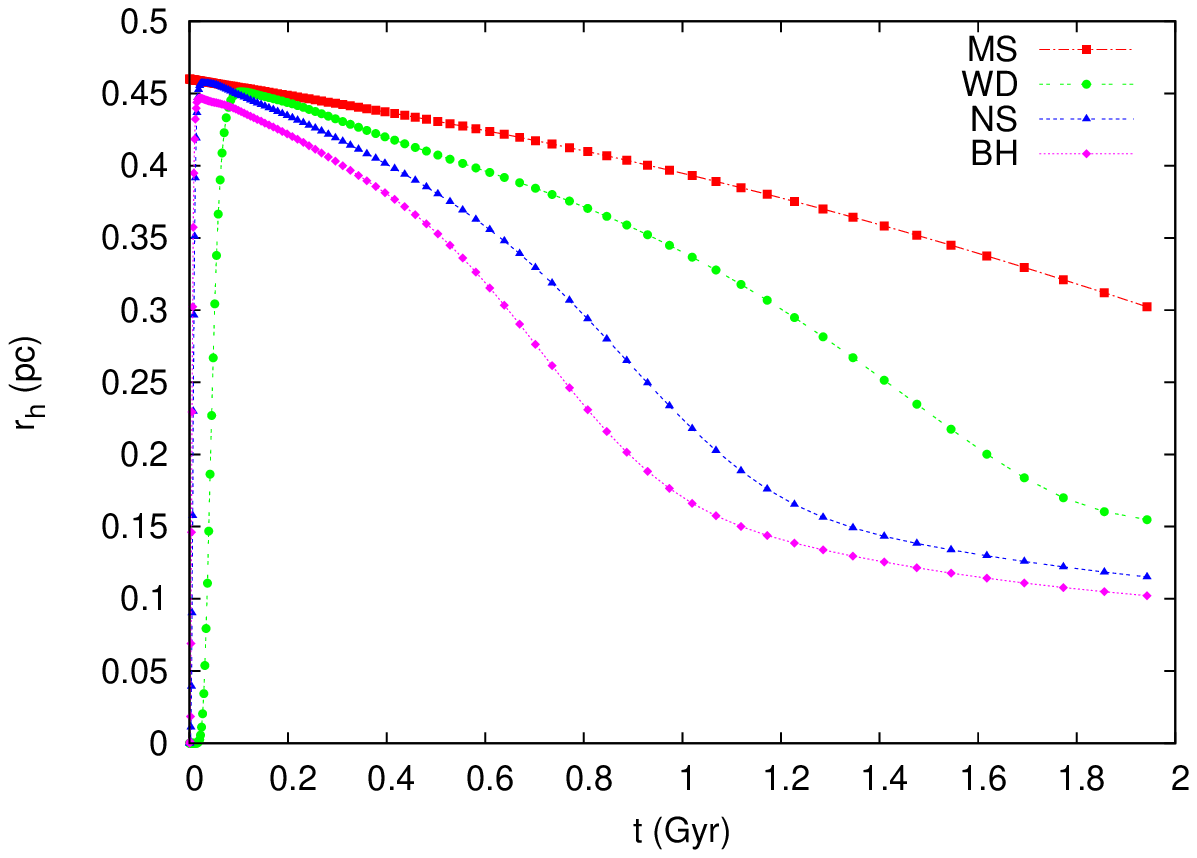}\\
\includegraphics[width=8cm]{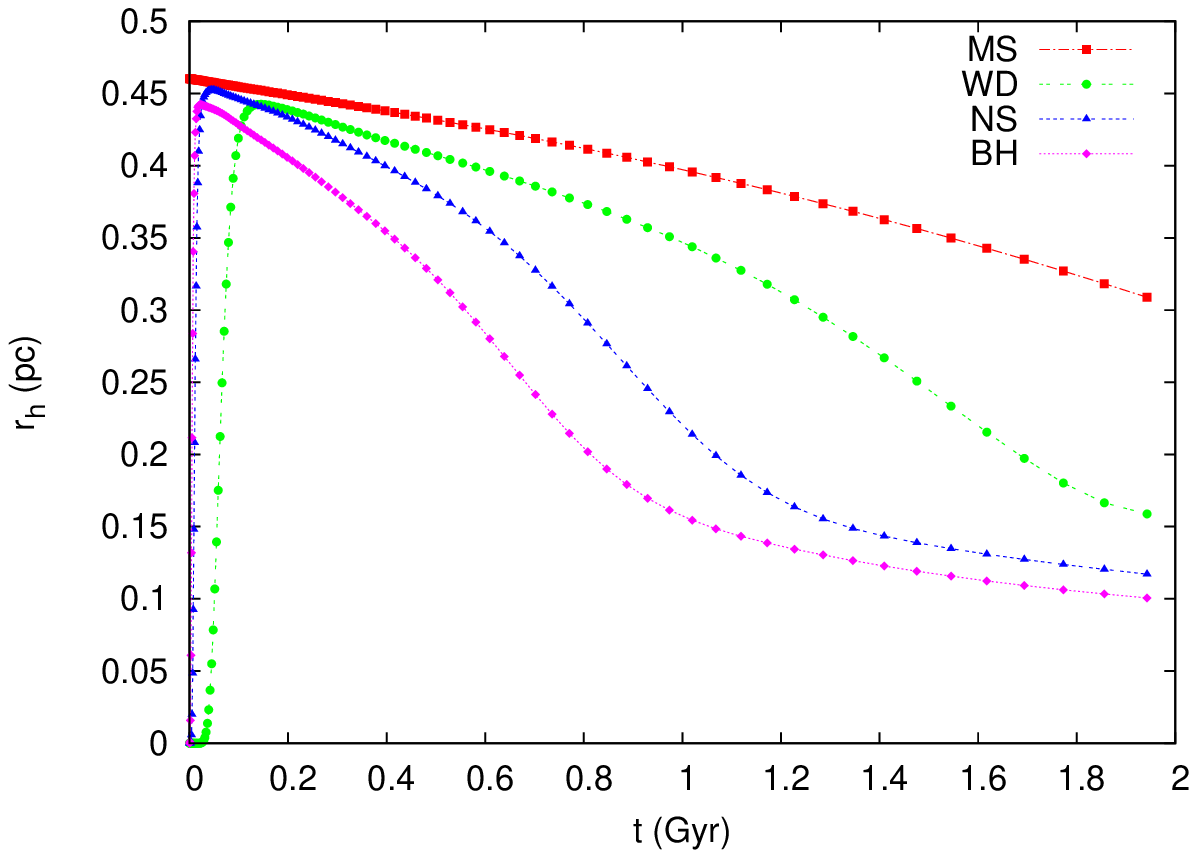}\\
\caption{Half-mass radius of different types of stars as a function of the time for a cluster with total mass $M_{\rm GC} = 10^6$ M$_\odot$. From top to bottom, panels refer to model A1, A2 and B2, respectively.}
\label{IMFevo}
\end{figure}

Figure \ref{IMFevo} shows the time evolution of the half-mass radius $r_{\rm h}$ of main sequence stars (MSs), white dwarfs (WDs), neutron stars (NSs) and black holes (BHs) for a GC with $M=10^6$ M$_\odot$ in models A1, A2 and B2. The interval of time considered is comparable to the time-scale needed to build up an MSS, as shown in Figure \ref{Fn2}.
It is evident that in all the cases considered, the population of BHs and NSs tends to concentrate in an inner region of the host GC, while WDs and MSs stars move in an outer region, having their motion less affected by df.

\begin{figure}
\includegraphics[width=8cm]{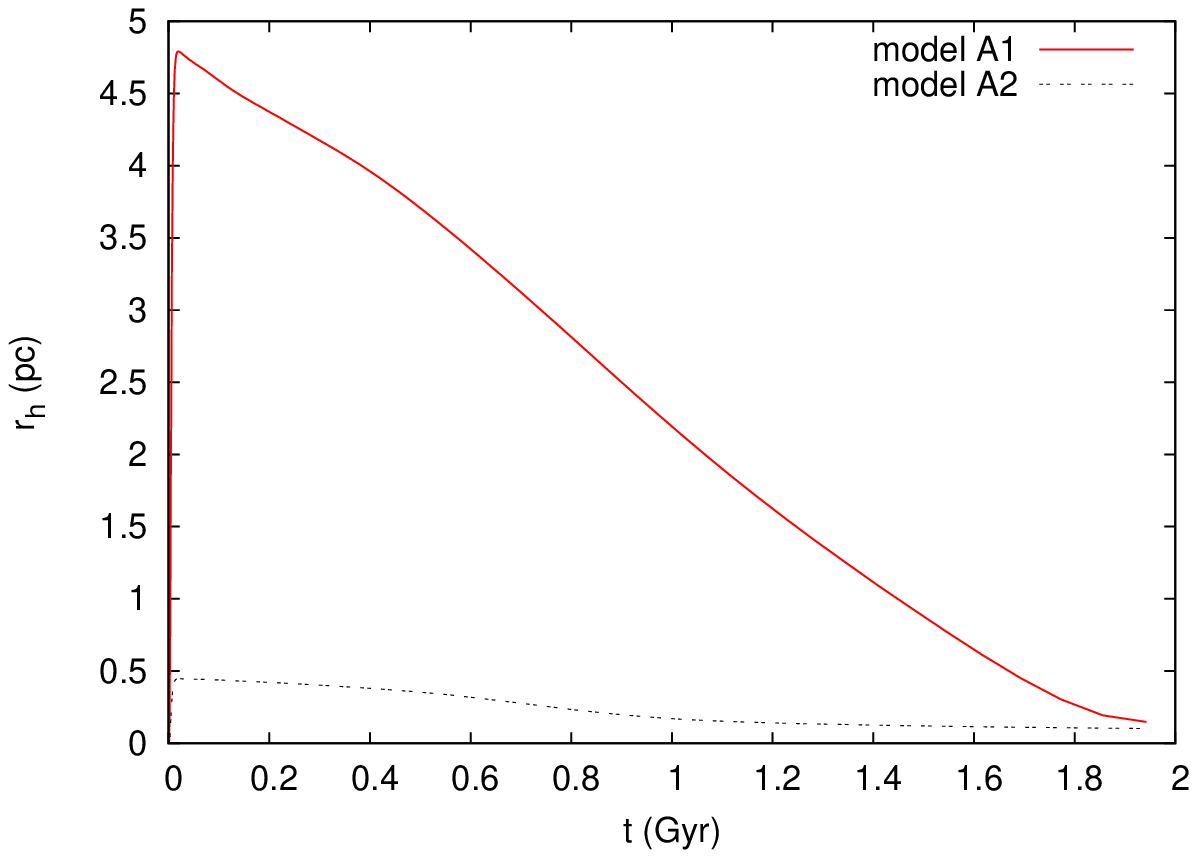}\\
\includegraphics[width=8cm]{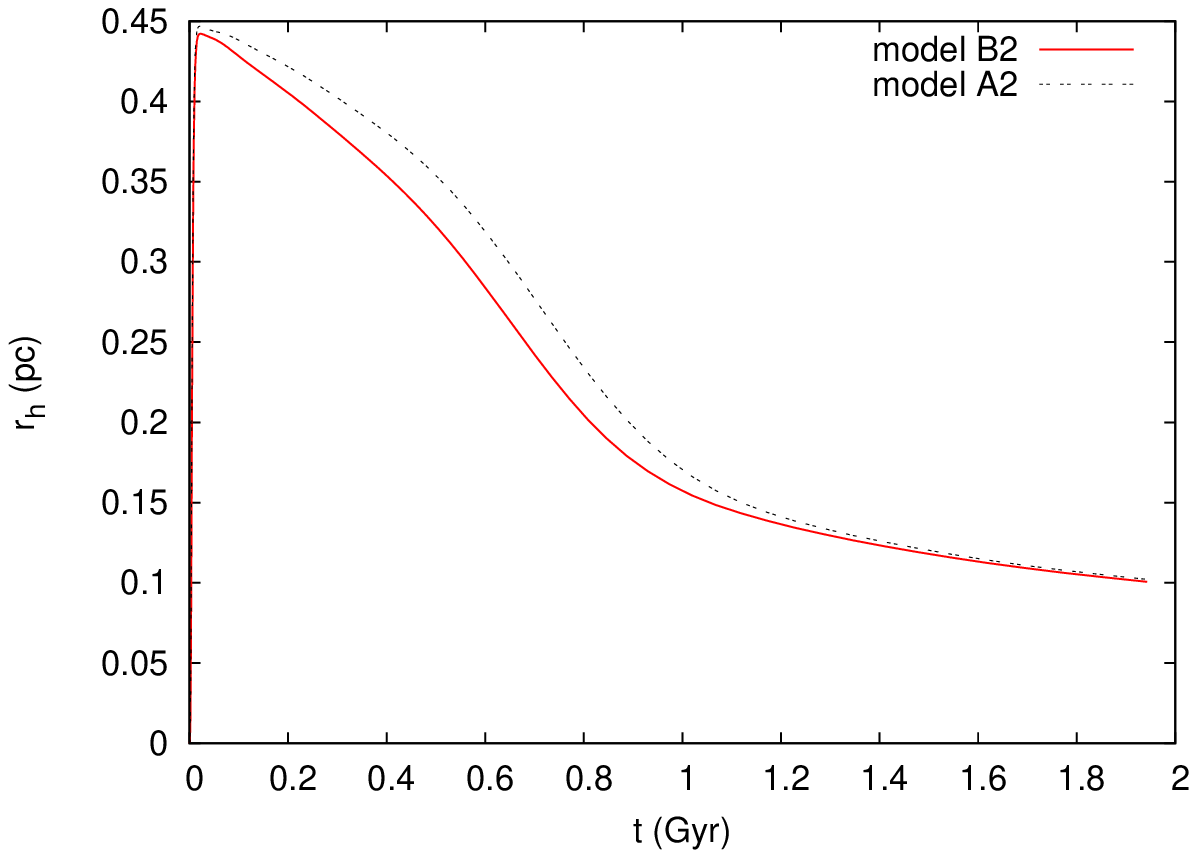}\\
\caption{Half-mass radius of BHs as a function of the time for a cluster with total mass $M_{\rm GC} = 10^6$ M$_\odot$. Top panel refers to models A1 (straight line) and A2 (dotted line) while bottom panel refers to models B2 (straight line) and A2 (dotted line), respectively.}
\label{IMFevo2}
\end{figure}

As shown in Figure \ref{IMFevo2}, the population of BHs concentrate more efficiently than other stellar types, reaching values $r_{\rm h,BH}\lesssim 0.1$ pc quite independently from the spatial distribution or the metallicity of the host cluster. 

Our results suggest that MSSs are mainly composed of BHs and NSs. On the other hand, it is currently quite ascertained that NSs can receive a kick at their birth caused by possible asymmetries that occur during the core-collapse of the NS progenitor \citep{gott70,harrison75,lai01}. 
Though the dispersion and distribution of the kick velocities is quite uncertain, most of the kicks estimates derived on theoretical and observational basis can throw the NS away from its host cluster (see for example \cite{podsi05} for a general discussion on this topic).  
The loss of NSs in the cluster can have significant effects on its global long term evolution, as shown recently by \cite{contenta15}.

Kick velocities are widely described through maxwellian distributions with $\sigma$ in the range $190-260$ km s$^{-1}$ \citep{hansen97,hobbs05}. 
However, this description seems to be at odds with observations, at least in part. 
Indeed, several GCs in the Milky Way contain more than $10^3$ NSs, whereas the kick velocities cited above should leave in the host cluster few tens of them. 
This problem, commonly referred to as ``NS retention problem'', can be partially solved if a substantial fraction of NSs form in massive binaries, where the kick momentum suffered by the NSs at their birth is shared with the companion, lowering the net velocity of the system \citep{drukier96,ivanova05,dalessandro11}. Some authors proposed that some NSs receive low birth kicks ($< 50$ km s$^{-1}$) depending on the properties of the binary in which they form \citep{pfahl02,podsi04}, or depending on the amount of mass that fall back during the core collapse of the star \citep{fryer12}.

In order to take in account which effect natal kicks can have on the population of NSs that compose the MSSs, we estimate in the following the fraction of NSs that are ejected in our GC models assuming that the velocity kicks are described by a maxwellian with $\sigma = 190$ km s$^{-1}$ as suggested by \cite{hansen97}.
In particular, we restrict this study applying natal kicks to models characterised by a Kroupa distribution of stellar masses and a Dehnen density profile. 

It is trivial to show that the local escape velocity from a given position $r_0$ is defined as
\begin{equation}
v_{\rm esc}(r_0)=\left(2\phi(r_0)\right)^{1/2},
\end{equation}
where $\phi(r_0)$ represents the gravitational potential generated by the cluster evaluated at the position $r_0$. 
Once the NS is kicked out, dynamical friction erases part of the kinetic energy excess generated by the kick. 
Indicating the energy depleted through df as $E_{\rm df}(r_0)$, the correct escape velocity is given by:
\begin{equation}
v_{\rm cor}(r_0) = v_{\rm esc}(r_0)\left(1+\frac{E_{\rm df}(r_0)}{\phi(r_0)}\right)^{1/2}.
\label{escape}
\end{equation}

Since $\phi(0)=(GM_{\rm GC}/r_{\rm GC})$ for a Dehnen model, the escape velocity from the innermost region of a GC with $M_{\rm GC} = 10^6$ M$_\odot$ and $r_{\rm GC}=0.15$ pc, is $v_{\rm esc}(0) \sim 170$ km s$^{-1}$. 

Following the numerical approach described by \cite{ASCD14df}, we found that the term $E_{\rm df}(r_0)/\phi(r_0)$ never exceeds $10^{-2}$ for a typical kick velocity in the range $10^2-10^3$ km s$^{-1}$, thus implying an increase of the escape velocity at most of $1\%$.

Therefore, assuming a maxwellian distribution of the kicks with $\sigma = 190$ km s$^{-1}$ and an escape velocity of $\sim 170$ km s$^{-1}$, we found that $\sim 10\%$ of the NSs that compose the MSS should remain bound to the host cluster, and therefore can come back to the MSS over a df time-scale.
However, this fraction can be much lower, depending on the GC properties and the dispersion of the velocity kicks, as shown in Table \ref{esc}.

\begin{table}
\caption{}
\centering{Escape velocity for a $M_{\rm GC}=10^6$ M$_\odot$.}
\begin{center}
\begin{tabular}{cccc}
\hline
\hline
$r_{\rm GC}$ & $v_{\rm cor}(0)$ & $f_{260} $ & $f_{190}$ \\
pc & km s$^{-1}$ & $\%$ & $\%$\\
\hline 
0.1 & 230 & 15  & 31 \\
0.2 & 148 & 4.5 & 10.5\\
0.5 & 93  & 1.2 & 2.9\\
1.0 & 66  & 0.4 & 1\\
\hline
\end{tabular}
\label{esc}
\end{center}
\begin{tablenotes}
\item Column 1: scale radius of the GC. Column 2: escape velocity evaluated through Equation \ref{escape}. Column 3: percentage of retained NSs assuming $\sigma=260$ kms$^{-1}$. Column 4: percentage of retained NSs assuming $\sigma=190$ kms$^{-1}$.
\end{tablenotes}
\end{table}

\subsection{Scaling relations}
\label{Scor}
In this section we provide scaling relations that link the MSS mass and the host GC mass.

As database to compare with our results we used the observational mass estimates for 14 putative IMBHs provided by LU13. In particular, LU13 have shown that the mass of the IMBH candidates is connected to the host cluster mass through the relation
\begin{equation}
{\rm Log} \left(\frac{M_{\rm IMBH}}{{\rm M}_\odot}\right) = (1.01\pm 0.34){\rm Log} \left(\frac{M_{\rm GC}}{{\rm M}_\odot}\right) - (2.37\pm 0.34).
\label{LU13}
\end{equation}
However, it should be noted that this relation is characterised by a relative error on its fitting parameters that exceeds $30\%$, thus implying quite large errors on the evaluation of the IMBH mass for a given mass of the host cluster.
Such errors are mostly determined by the relatively small amount of data available and the large errors that characterise them. Hence, here we compare only the statistical behaviour of the scaling relation, whereas a more robust comparison need a larger amount of data and a much higher precision in the evaluation of the mass excess.

\begin{figure*}
\centering
\includegraphics[width=8cm]{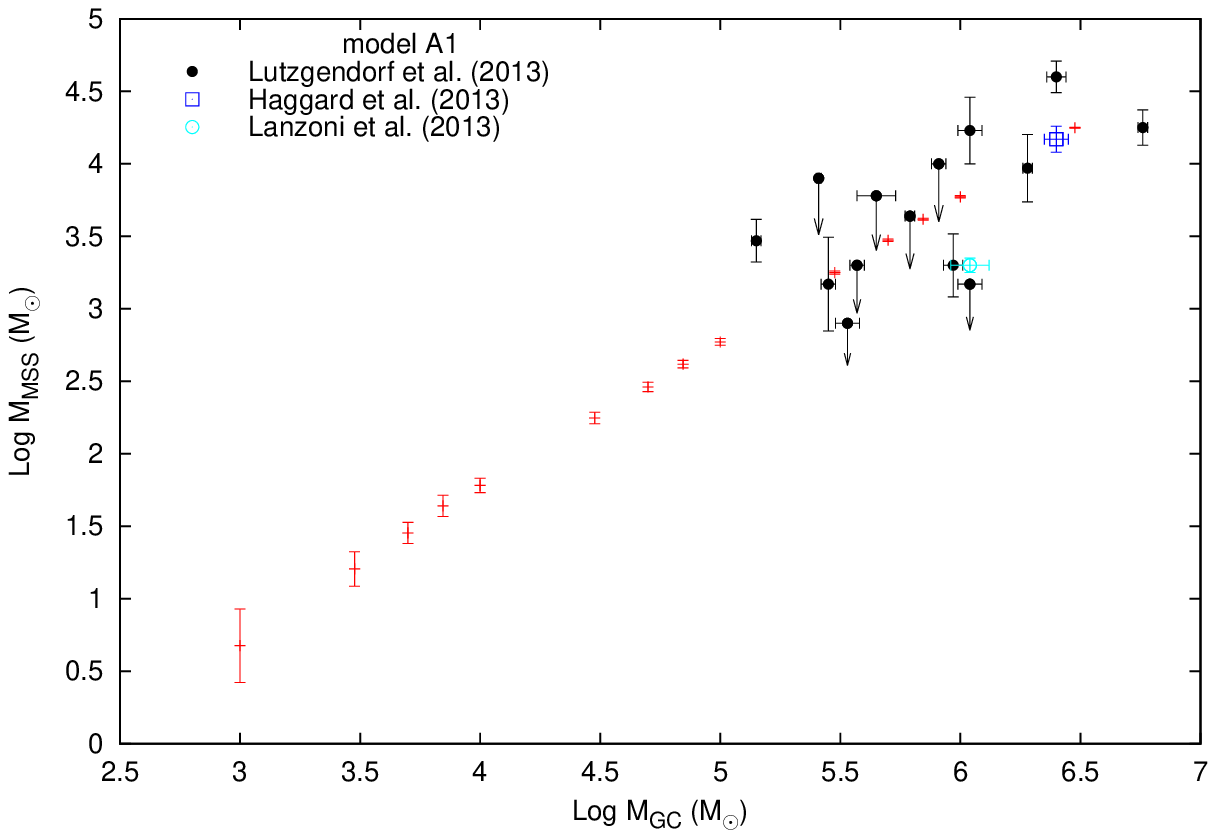}
\includegraphics[width=8cm]{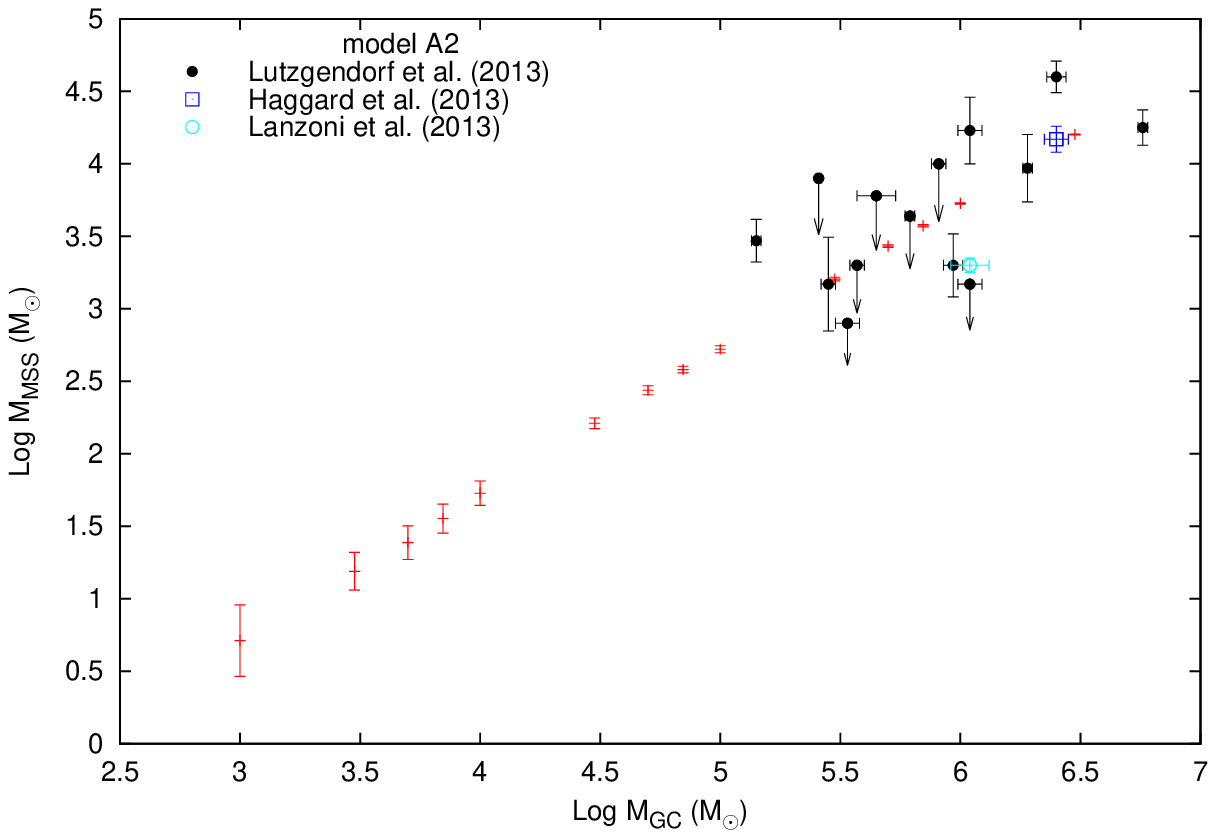}\\
\includegraphics[width=8cm]{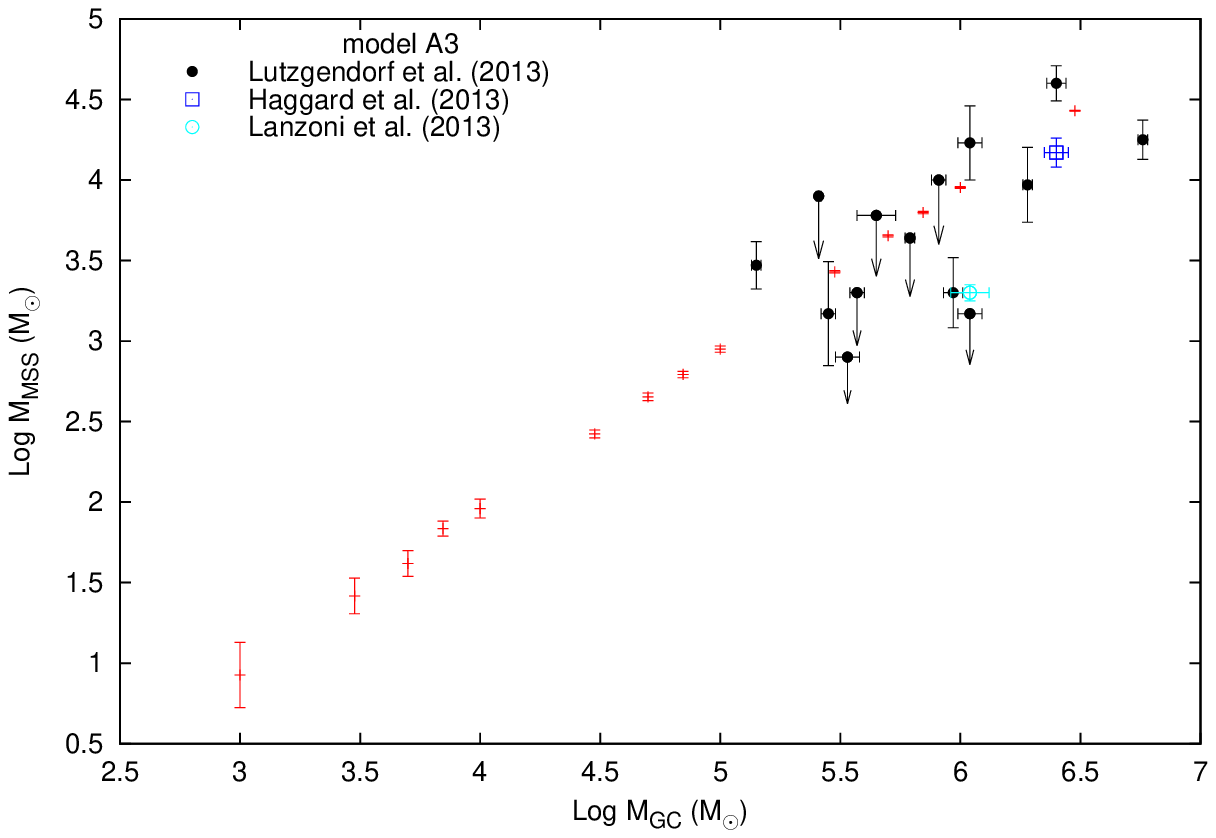}
\includegraphics[width=8cm]{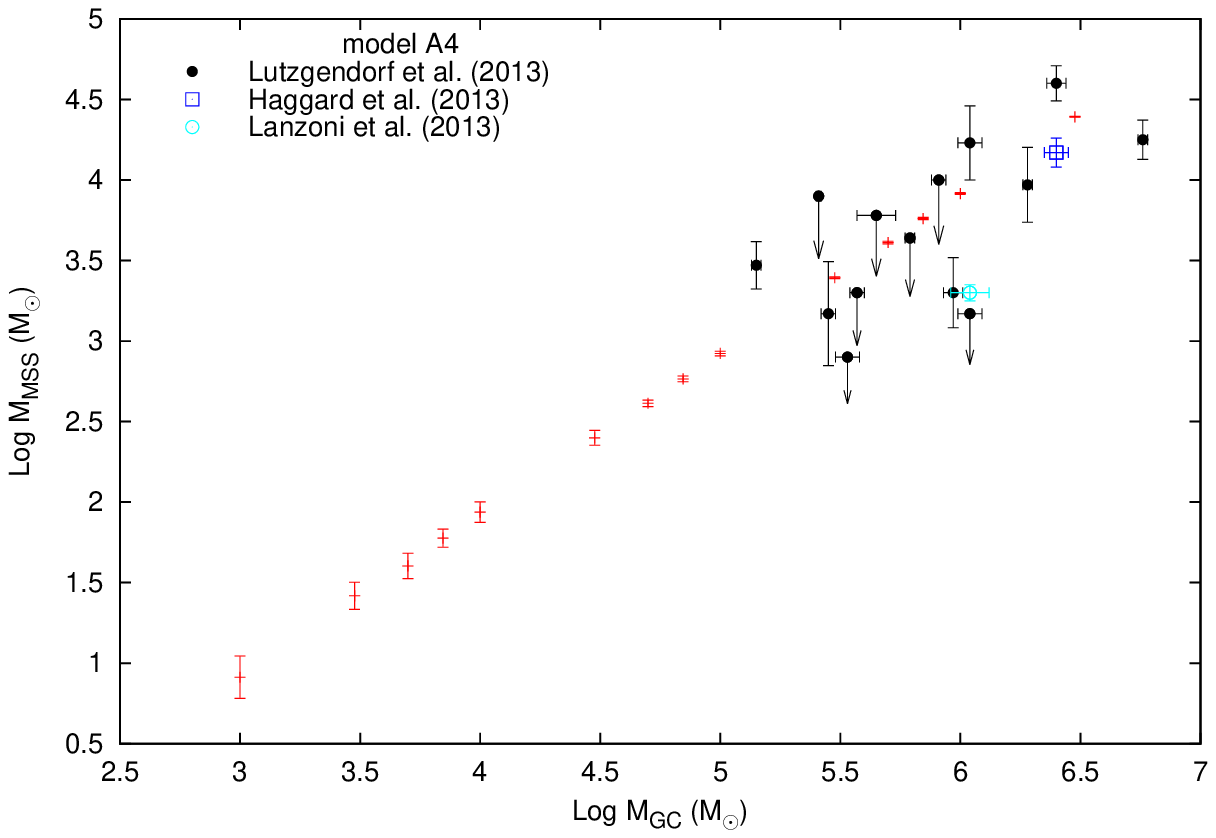}\\
\includegraphics[width=8cm]{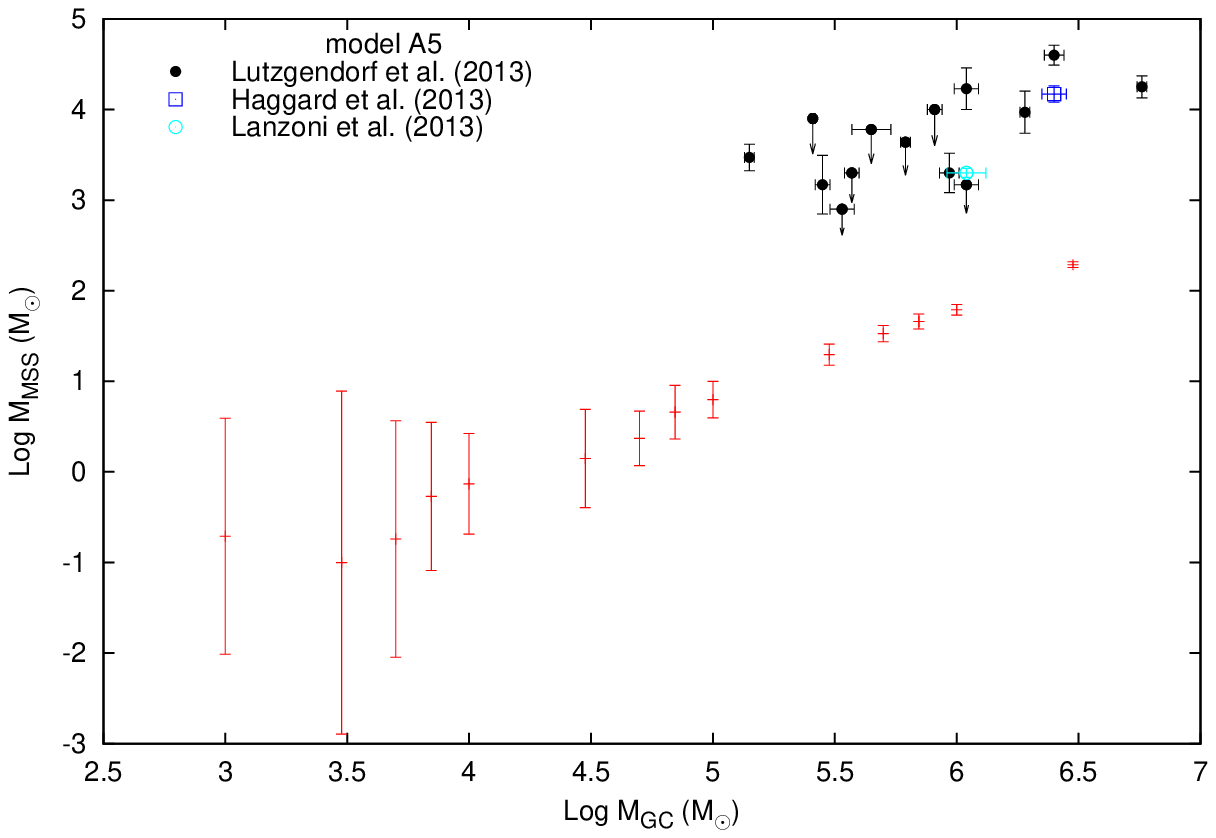}
\includegraphics[width=8cm]{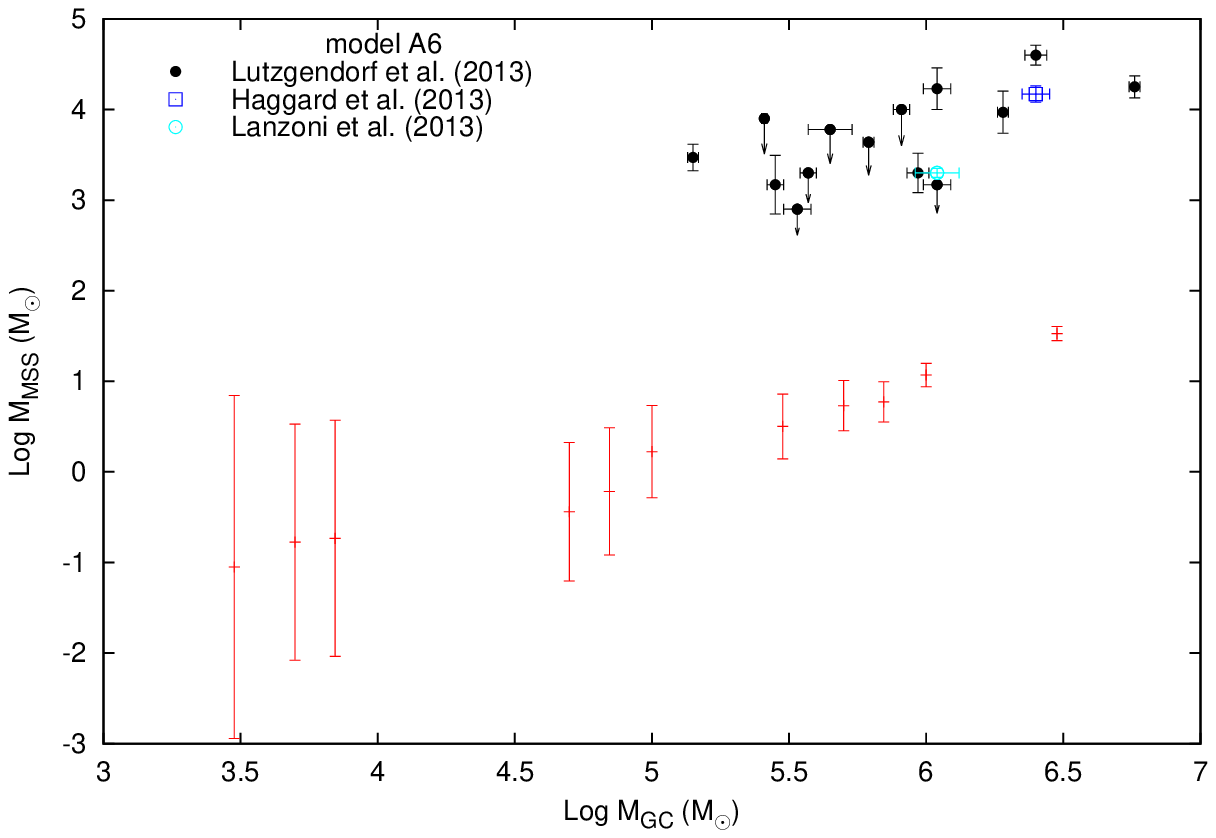}
\caption{Segregated mass vs. the total mass of the cluster. Black filled circles represent LU13 data. The blue open square indicates the central mass estimates for the cluster $\omega$ Cen provided by \citet{haggard13}, while the cyan open circle indicates the central mass hosted in the cluster NGC6388 as suggested by \citet{lanzoni13}. All the models in this case have solar metallicities.}
\label{F5}
\end{figure*}

\begin{figure*}
\centering
\includegraphics[width=8cm]{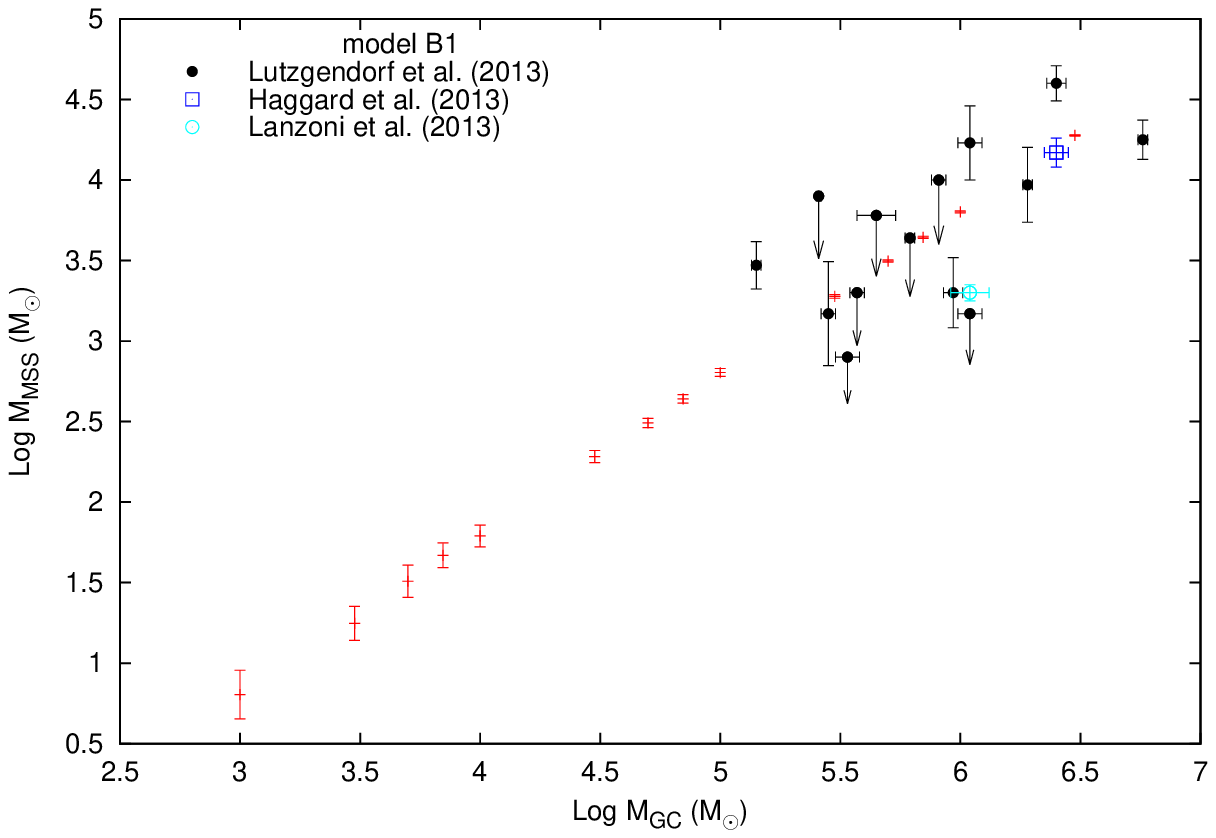}
\includegraphics[width=8cm]{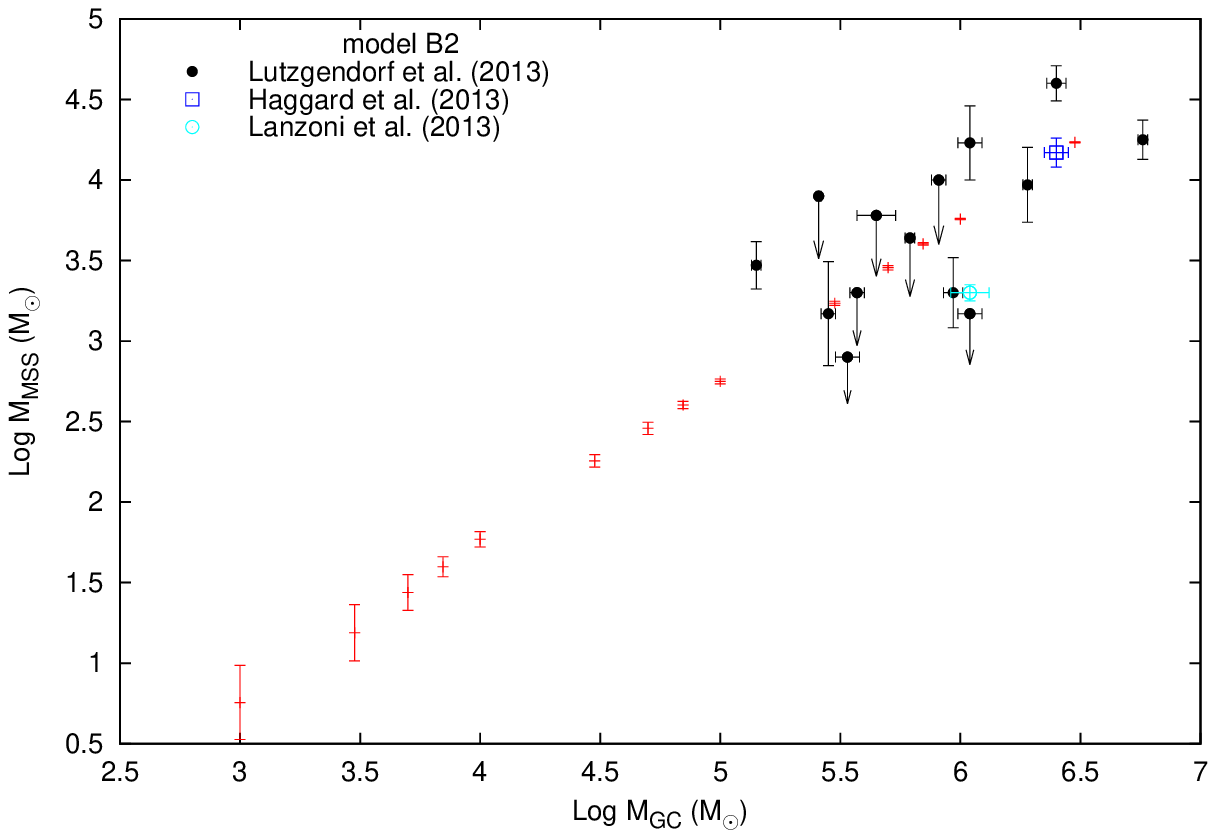}\\
\includegraphics[width=8cm]{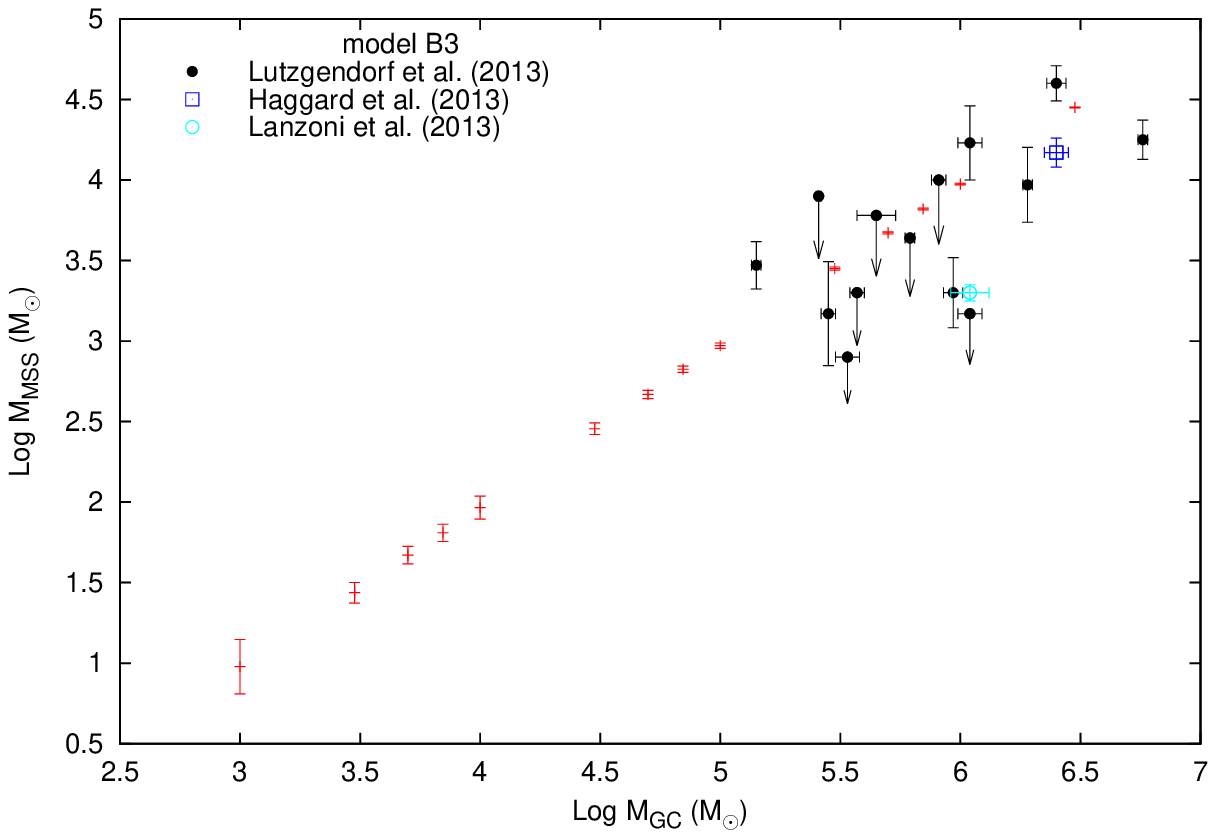}
\includegraphics[width=8cm]{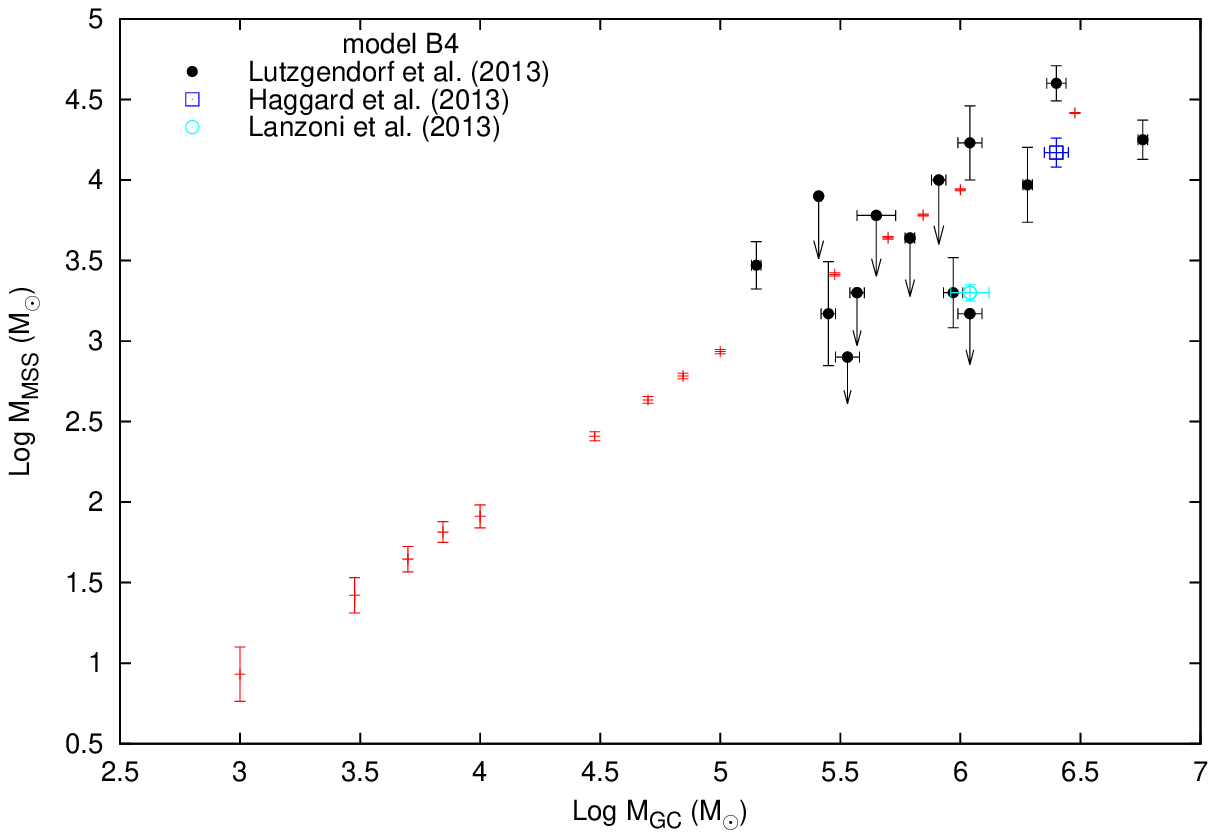}\\
\includegraphics[width=8cm]{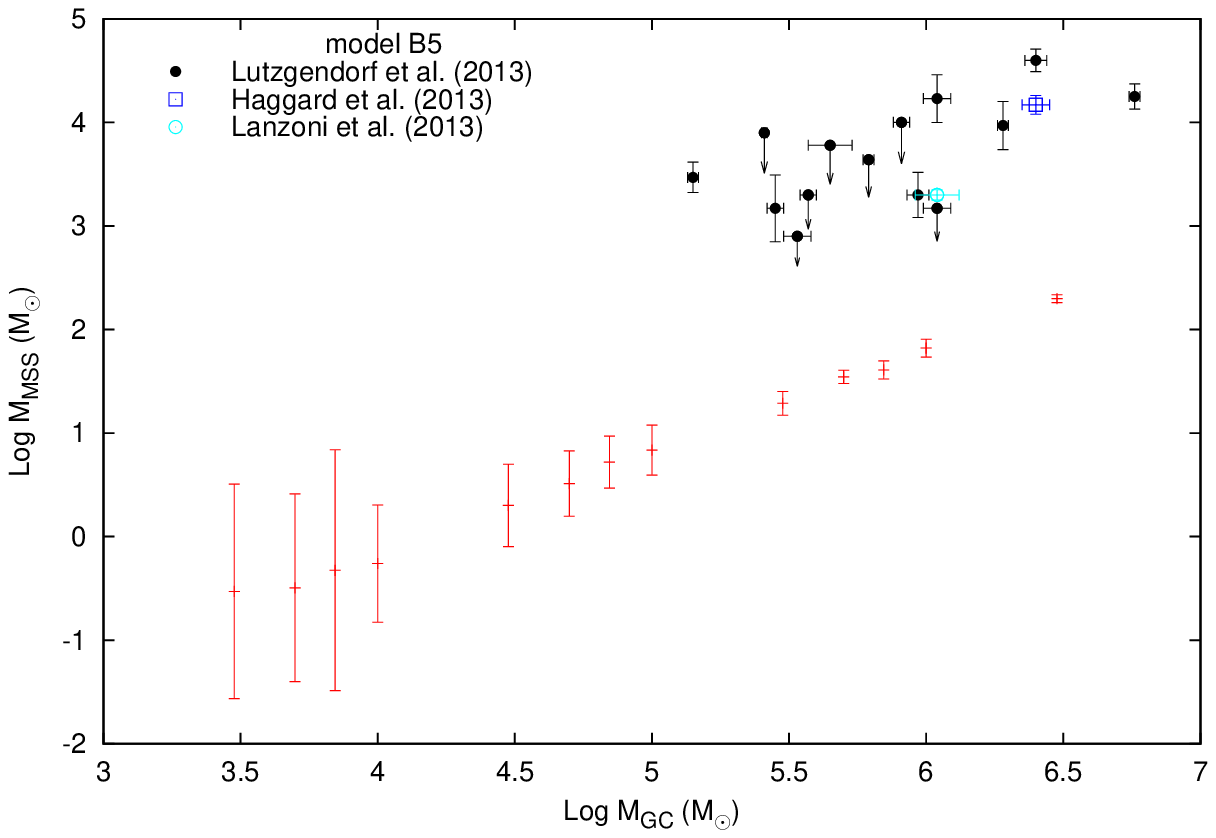}
\includegraphics[width=8cm]{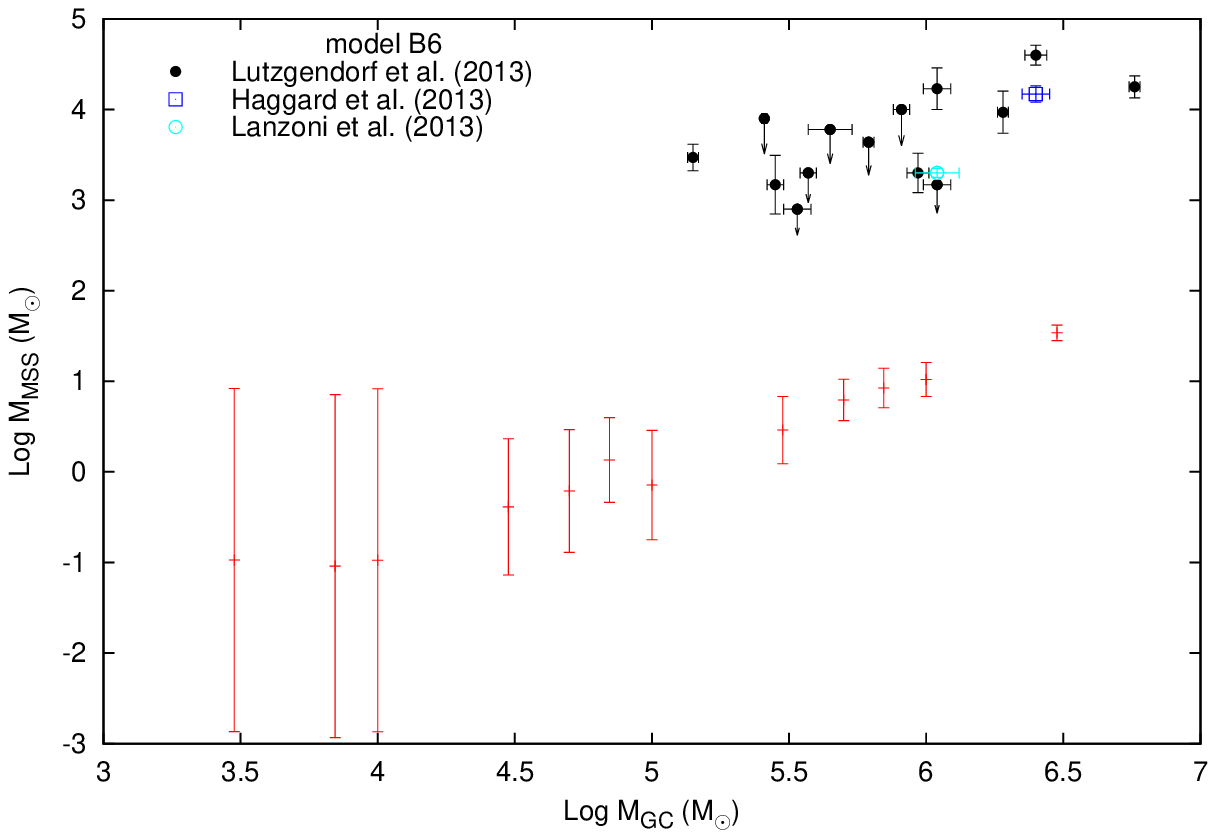}
\caption{As in Figure \ref{F5}, but here all the models have a value of the metallicity $Z=0.0004$.}
\label{F6}
\end{figure*}

Looking at Figures \ref{F5} and \ref{F6}, which show the MSS masses versus the initial mass of the host GCs for all the cases considered, it is evident that the best agreement with the observational estimates provided by LU13 is achieved for models characterised by a Kroupa IMF, labelled with numbers 1 and 2, and a Salpeter IMF, labelled with numbers 3 and 4.

As expected, models with a flat IMF, labelled with number 5 and 6, produce MSSs with masses that are clearly much smaller than the observational estimates.
 
Nevertheless, a flat IMF is completely at odds with observations, and therefore it is not surprising to find MSS masses very different from observational estimates in these models.

For all the cases investigated here, we found the following power-law best-fitting formula:
\begin{equation}
{\rm Log} \left(\frac{M_{\rm MSS}}{{\rm M}_\odot}\right) = a{\rm Log} \left(\frac{M_{\rm GC}}{{\rm M}_\odot}\right) + b,
\label{eqco}
\end{equation}
where the slope, $a$, and the offset, $b$, of the correlation were computed by means of a Marquardt-Levenberg non-linear regression algorithm. The values of $a$ and $b$ are summarised in Table \ref{corr1}.

It is worth noting that different values of $Z$ produce small differences in the correlations. In particular, we found that models with low metallicity produces MSSs with masses $5\%$ greater than that obtained for models with solar metallicities, thus implying at most a difference smaller than $1\%$ on the determination of the parameters of the best-fitting formula.

Comparing our best-fitting parameters with that provided in LU13, it is evident an overall agreement, despite the LU13 results are affected by uncertainties that in some cases exceed $2$ dex.

Some estimates contained in the LU13 dataset seem to be at odds with other works. For instance, some observations suggest that the cluster $\omega$ Cen hosts in its innermost region a mass in the range $1.2-1.8 \times 10^4$ M$_\odot$ (see for example \cite{vandermarel10} and \cite{haggard13}), less than a half with respect to the value suggested by LU13. 
Another interesting case is the cluster NGC6388, for which \cite{lanzoni13} suggested a central mass of $2\times 10^3$ M$_\odot$, whereas LU13 found a greater value, $\sim 1.7\times 10^4$.
Using the scaling relations drawn here, we can give an estimate of the central mass hosted in these two systems.
To model the old, massive cluster $\omega$ Cen, we assume a low value of the metallicity, a Kroupa IMF and a cored density profile. Note that this choice corresponds to models labelled with B2. Using the corresponding parameters (see Table \ref{corr1}) and assuming a total mass $M_{\rm GC} = 2.5\times 10^6$ M$_\odot$ (see Table \ref{TABGC}), we found $M_{\rm MSS} = (1.45\pm 0.03) \times 10^3$ M$_\odot$, in agreement with the values provided by \cite{vandermarel10} and \cite{haggard13}. Applying the same procedure to the cluster NGC6388, we found $M_{\rm MSS} \simeq (6.4\pm 1.2)\times 10^3$ M$_\odot$, a value which lies between the estimates provided by LU13 and \cite{lanzoni13}. A numerical study focused on a detailed modelling of this cluster may help to shed light on the possible causes of these discrepancies \citep{LU15}.

Finally, for the sake of comparison, we investigated whether our results agree with previous theoretical works.  
For instance, using Monte Carlo simulations, \cite{GURK} have shown that the mass of the shrinking core that forms in GCs as a consequence of mass segregation, is linked to the cluster mass through the relation:
\begin{equation}
{\rm Log} \left(\frac{M_{\rm MSS}}{{\rm M}_\odot}\right) = {\rm Log} \left(\frac{M_{\rm GC}}{{\rm M}_\odot}\right) - (2.8\pm 0.2),
\end{equation}
in quite good agreement with our estimates both for solar and low metallicities.

\cite{zwart02}, instead, found that the maximum mass of an IMBH formed through runaway collisions in a GC with a mass above $10^6$ M$_\odot$ is given by:
\begin{equation}
{\rm Log} \left(\frac{M_{\rm IMBH}}{{\rm M}_\odot}\right) = {\rm Log} \left(\frac{M_{\rm GC}}{{\rm M}_\odot}\right) - 2.1.
\label{spzscaling}
\end{equation}

The similarity between our scaling relations and Equation \ref{spzscaling} confirm that MSSs represent the ideal reservoir of stars which can contribute to the assembly of an IMBH in a runaway fashion.

\begin{table}
\caption{}
\centering{Slope of the scaling relation between MSS masses and the host cluster masses.}
\begin{center}
\begin{tabular}{ccccc}
\hline
\hline
Model name & $a$ & $\epsilon_a$ & $b$ & $\epsilon_b$ \\
\hline
A1 & $1.000$ & $0.002$ & $-2.23$ & $0.01$ \\
A2 & $0.996$ & $0.002$ & $-2.25$ & $0.01$ \\
A3 & $0.999$ & $0.001$ & $-2.045$ & $0.008$\\
A4 & $0.998$ & $0.001$ & $-2.075$ & $0.006$\\
A5 & $1.01$ & $0.01$ & $-4.233$ & $0.082$\\
A6 & $0.98$ & $0.04$ & $-4.86$ & $0.27$\\
B1 & $1.000$ & $0.001$ & $-2.202$ & $0.007$\\
B2 & $0.999$ & $0.001$ & $-2.238$ & $0.009$\\
B3 & $0.999$ & $0.001$ & $-2.024$ & $0.005$\\
B4 & $1.001$ & $0.001$ & $-2.066$ & $0.007$\\
B5 & $0.99$ & $0.01$ & $-4.160$ & $0.087$\\
B6 & $0.97$ & $0.03$ & $-4.77$ & $0.20$\\
\hline
\end{tabular}
\end{center}
\begin{tablenotes}
\item Column 1: model name. Column 2-3: slope of the fitting function and relative error. Column 4-5: offset of the fitting function and relative error. 
\end{tablenotes}
\label{corr1}
\end{table}

\section{$N$-body modelling of a stellar cluster: IMF, mass segregation and MSS early formation}
\label{nbody}
 
In order to compare with our statistical results, we ran 10 direct $N$-body simulations of star clusters characterised by broad IMFs and masses up to $3\times 10^5$ M$_\odot$. In particular, these models consist of $8-16-32-262-524$k particles. 

To perform the simulations, we used the HiGPUs code \citep{Spera}, a direct $N$-body integrator which fully exploits the advantages arising from parallel computing. 
HiGPUs allows very fast integration keeping a very high level of precision but, on the other hand, it does not implement any treatment of binary formation and close encounters and hence, it does not allow to follow the long-term evolution of the system. 

Moreover, the current version of HiGPUs does not contain any recipe for stellar evolution. To improve the comparison with our semi-statistical results, we developed a modified version of HiGPUs in which we included the SSE package.
In the following, we will refer to this modified version as HiGPUsSE. We labelled in the following with SEn the simulations performed with HiGPUs, whereas we used SEy to label those carried out with HiGPUsSE.

We set as gravitational softening $\epsilon = 0.05$ pc, thus allowing a reliable description of the evolution of the system as long as $\epsilon$ is sufficiently smaller than the mean inter-particle distance, $\lambda$, defined as \citep{gilbert68,boily99,nelson99}:
\begin{equation}
\lambda = F\frac{1}{n^{1/3}},
\label{lamb}
\end{equation}
where $F\sim 0.9$ and $n$ is the numerical density given by the stars which move in the innermost region of the cluster. 

A good estimate of $\lambda$ is provided by the lagrangian radius which contains the $0.1\%$ of the total GC mass. Hence, in the following we will use this typical radius to check whether our results provide a proper description of the dynamical evolution of the system.

Since our methodology is not well suited to describe the long-term evolution of the cluster nucleus, we integrated these models up to $\simeq 10^2$ crossing times, a time-scale sufficiently shorter than the relaxation time, but long enough to highlight and characterise the formation of an MSS.

Stars' masses are sampled according to a Salpeter or a Kroupa IMF. 
Moreover, to model the clusters we used either a Dehnen density profile or an uniform density profile, as we have done for the statistical work discussed above. 
As shown in Table \ref{Nbody}, each model is labelled with the number of particles used to represent it, the kind of code used to evaluate dynamics and stellar evolution, the IMF used to sample stellar masses and the density distribution used to sample stars' positions.

\begin{table*}
\caption{}
\centering{Parameters of the $N$-body simulations.}
\begin{center}
\begin{tabular}{lccccccccc}
\hline
\hline
Model ID & code & IMF & $\rho(r)$ & $\gamma$ & $r_{\rm GC} $ & $R_{\rm GC}$ & $M_{\infty} $ & $M_{\rm GC} $  & $N$ \\
 & & & & & (pc) & (pc) & ($10^3$ M$_\odot$)& ($10^3$ M$_\odot$)& \\
\hline
\texttt{8k\_SEn\_U\_S} & HiGPUs & S    & U & 0 & 5 & 5 & 0.3& 0.3& 8024 \\
\texttt{8k\_SEy\_U\_S}& HiGPUsSE & S  & U & 0 & 5 & 5 & 0.3& 0.3& 8024 \\
\texttt{16k\_SEn\_U\_S}& HiGPUs & S    & U & 0 & 5 & 5 & 5.1& 5.1& 16384 \\
\texttt{16k\_SEy\_U\_S}& HiGPUsSE & S  & U & 0 & 5 & 5 & 5.1& 5.1& 16384 \\
\texttt{32k\_SEn\_U\_S}& HiGPUs & S    & U & 0 & 5 & 5 & 11&  11& 32768 \\
\texttt{32k\_SEy\_U\_S}& HiGPUsSE & S  & U & 0 & 5 & 5 & 11&  11& 32768 \\
\texttt{262k\_SEn\_D\_S}& HiGPUs & S   & D & 0 & 0.7 & 10 &100&  90& 262144  \\
\texttt{262k\_SEy\_D\_S}& HiGPUsSE & S & D & 0 & 0.7 & 10 &100&  90& 262144 \\
\texttt{524k\_SEn\_D\_K}& HiGPUs & K   & D & 0 & 1.6 & 10 &500& 335& 524288 \\
\texttt{524k\_SEy\_D\_K}& HiGPUsSE & K & D & 0 & 1.6 & 10 &500& 335& 524288 \\
\hline
\label{Nbody}
\end{tabular}
\end{center}
\begin{tablenotes}
\item Column 1: model name. Column 2: $N$-body code used. Column 3: IMF used to sample the mass of each star: Salpeter (S) or Kroupa (K) mass function. Column 4: density profile used to sample stars' positions: uniform (U) or Dehnen (D) density distribution. Column 5: slope of the density profile. Column 6: scale radius of the model in pc. Column 7: total radius of the cluster. Column 8: mass of the cluster for $r_{cut}\rightarrow \infty$. Column 9: mass of the cluster actually sampled. Column 10: number of particles used to represent the cluster.
\end{tablenotes}
\end{table*}

\subsubsection{Dynamical friction}
Direct $N$-body simulations are powerful tools to follow the mass segregation of stars in stellar clusters, since in this case df comes out naturally as a consequence of the two-body interactions.
On the other hand, the presence of a softening length that smooths the gravitational interaction may lead to less efficient energy exchanges among particles, altering in some cases the orbital decay process. 
Due to this reason, we ran several test simulations of the same model at decreasing values of $\epsilon$. We find that the orbital evolution of the most massive stars is substantially unaltered when $\epsilon<0.05$ pc. Hence, to obtain a good compromise between the computational cost and the accuracy of the simulations, we set this limiting value as softening length.

\begin{figure}
\centering
\includegraphics[width=8cm]{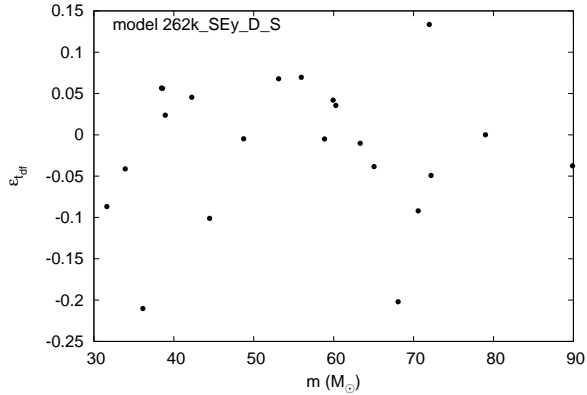}\\
\caption{Relative error, $\varepsilon_{\rm df} = 1-t_{\rm df}/t_{\rm dfNB}$, as a function of the star masses, between the df time-scale evaluated through Equation \ref{tdf} and through the $N$-body simulation in the case of model \texttt{262k\_SEy\_D\_S}.}
\label{traj}
\end{figure}

An accurate $N$-body modelling of the cluster would allow to get the dynamical friction time-scale of stars. Hence, we tested the validity of our treatment for df comparing the theoretical estimates of $t_{\rm df}$ with the values obtained directly from the simulations. In particular, we selected in simulation \texttt{262k\_SEy\_D\_S} those stars for which the orbital decay occurs over a time sufficiently greater than the time resolution of the simulation itself. This requirement is crucial in order to extrapolate the decay time of a stars in the $N$-body model. Having detailed informations about the orbital evolution of these stars, we evaluated their df times, $t_{\rm df,NB}$, comparing them with the theoretical estimates obtained through Equation \ref{eq2cor}.
In particular, Figure \ref{traj} shows the relative difference, $\varepsilon_{\rm df}$, between the $N$-body and the theoretical time-scale provided by Equation \ref{eq2cor}:
\begin{equation}
\varepsilon_{\rm df} = 1-t_{\rm df}/t_{\rm dfNB}.
\end{equation}

The maximum difference $\varepsilon_{\rm df}$ between semi-analytic treatment and N-body simulations is $\sim  20\%$. Extrapolating the orbital parameters of stars is the main source of uncertainty for $t_{\rm df,NB}$, due to the fact that our simulations have a poor time-resolution.

The good agreement between the semi-analytical estimates and the simulations suggest that Equation \ref{eq2cor} can be used to reproduce reliable sets of initial conditions for the cluster immediately after the completion of the mass segregation process, thus allowing to simulate only the subsequent evolution of the cluster.

It should be noted that a detailed simulation of the mass segregation process for a reliable cluster model with a mass above $10^5$ M$_\odot$ may require more than a month to be performed, even using the most advanced direct $N$-body codes and hardware facilities currently available. 
Hence, our approach can allow to gain a significant amount of time for simulating the GC evolution.

As the mass segregation proceeds, heavy stars concentrate toward the innermost region of the host cluster, leading to a progressive increase of the density, which in turn enhances the probability to have close encounters among stars.

In HiGPUs and HiGPUsSE, gravitational encounters are smoothed through the gravitational softening $\epsilon$.
Hence, we stopped our simulations when the lagrangian radius which contains $0.1\%$ of the total mass of the cluster became comparable to $\epsilon$, in order to reproduce in the most reliable way the early evolution of the host nucleus.

\begin{figure}
\centering
\includegraphics[width = 8cm]{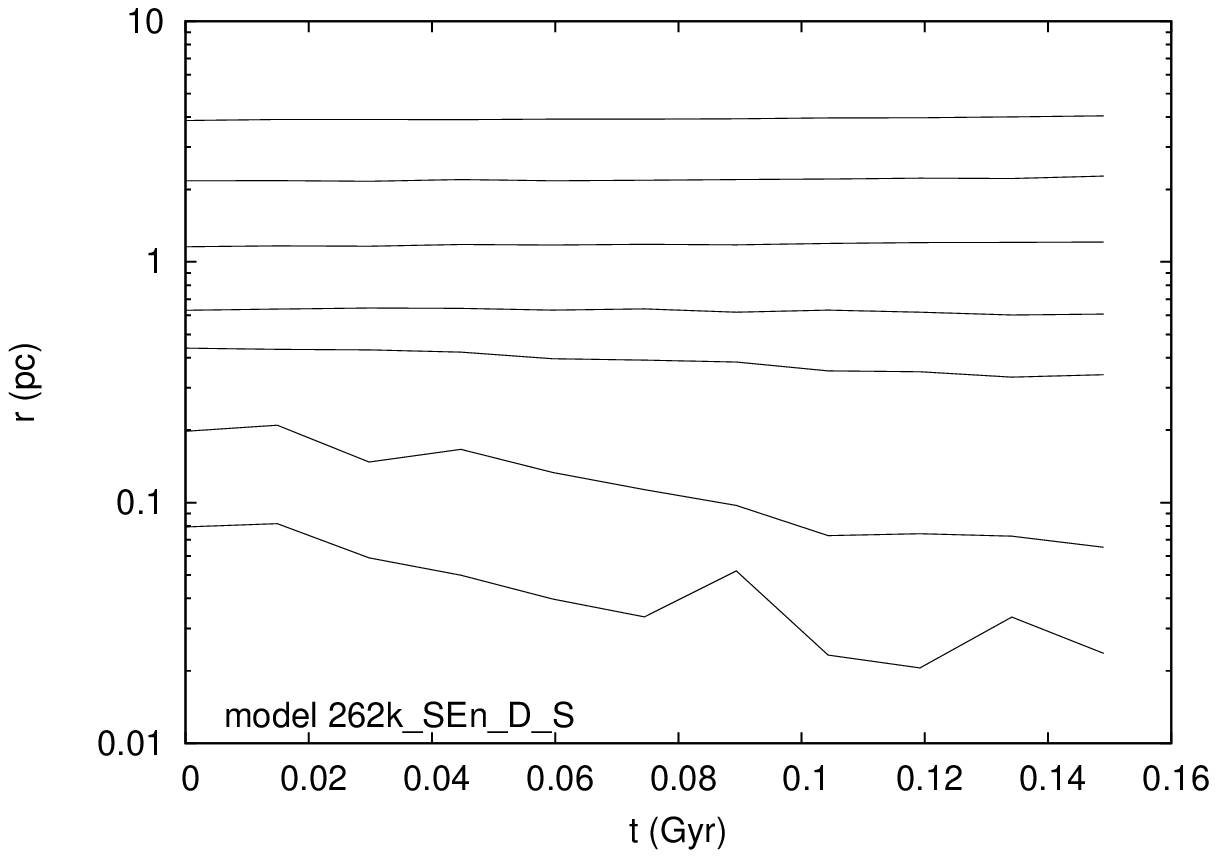}\\
\includegraphics[width = 8cm]{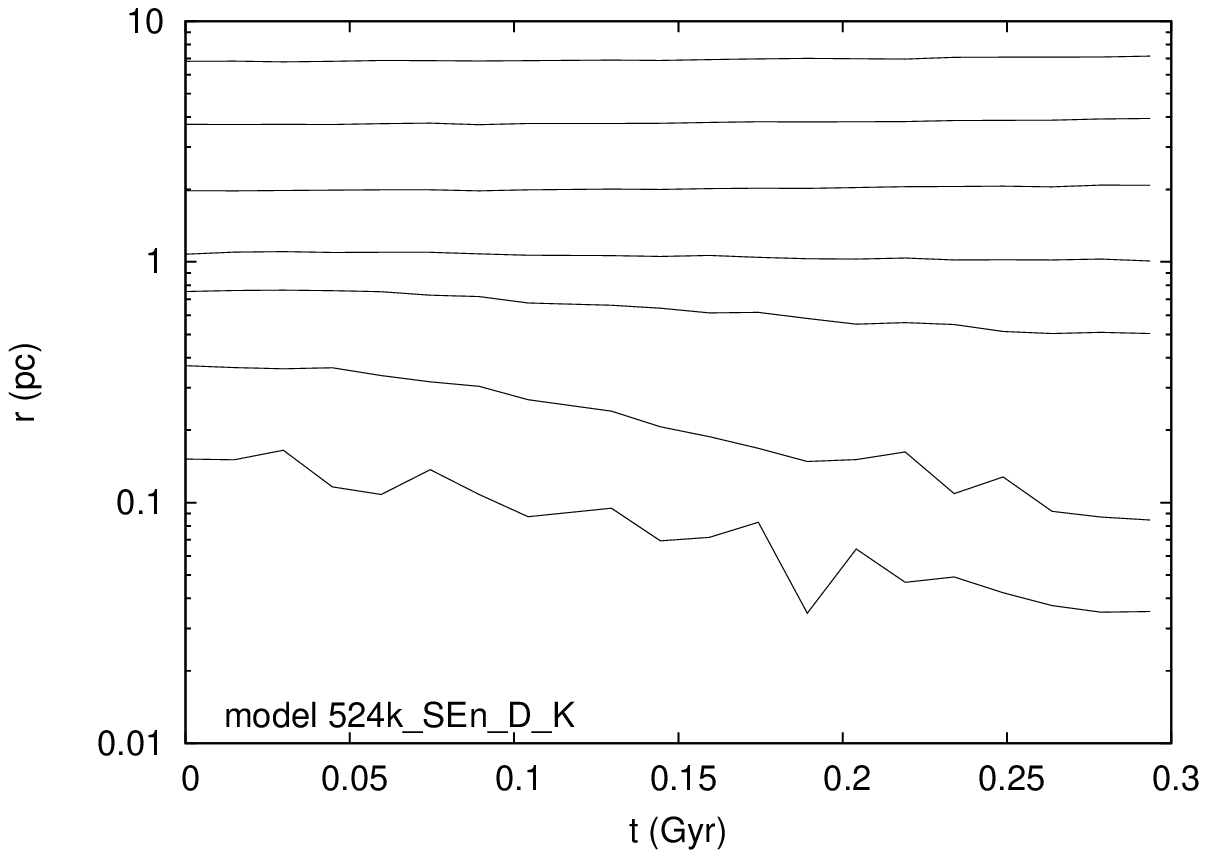}\\
\caption{Evolution of the lagrangian radii as a function of the time. Top panel refers to simulation \texttt{262k\_SEn\_D\_S}, whereas bottom panel refers to simulation \texttt{524k\_SEn\_D\_K}.
In each panel, from bottom to top, each line represents the lagrangian radius containing $0.1-1-5-10-25-50-75\%$ of the total mass of the cluster. }
\label{lagr}
\end{figure}

For instance, looking at the time evolution of lagrangian radii shown in Figure \ref{lagr} makes evident that the innermost lagrangian radius reaches a size comparable to the softening in $t \simeq 100$ Myr for simulation \texttt{262k\_SEn\_D\_S} and $t \gtrsim 200$ Myr for simulation \texttt{524k\_SEn\_D\_K}.

\subsubsection{Stellar evolution}

In this section we try to determine if stellar evolution can affect significantly the formation of an MSS. In order to do this, we will compare the simulations performed with HiGPUs with those obtained with its modified version, HiGPUsSE.
On the other hand, it should be stressed that at the time, nor HiGPUs neither HiGPUsSE include any treatment for close interactions and do not account for NSs' natal kicks. Hence, we limit our comparison to the time interval over which the MSS forms and without considering dynamical kicks suffered by NSs at their birth.

\begin{figure*}
\centering
\includegraphics[width = 8cm]{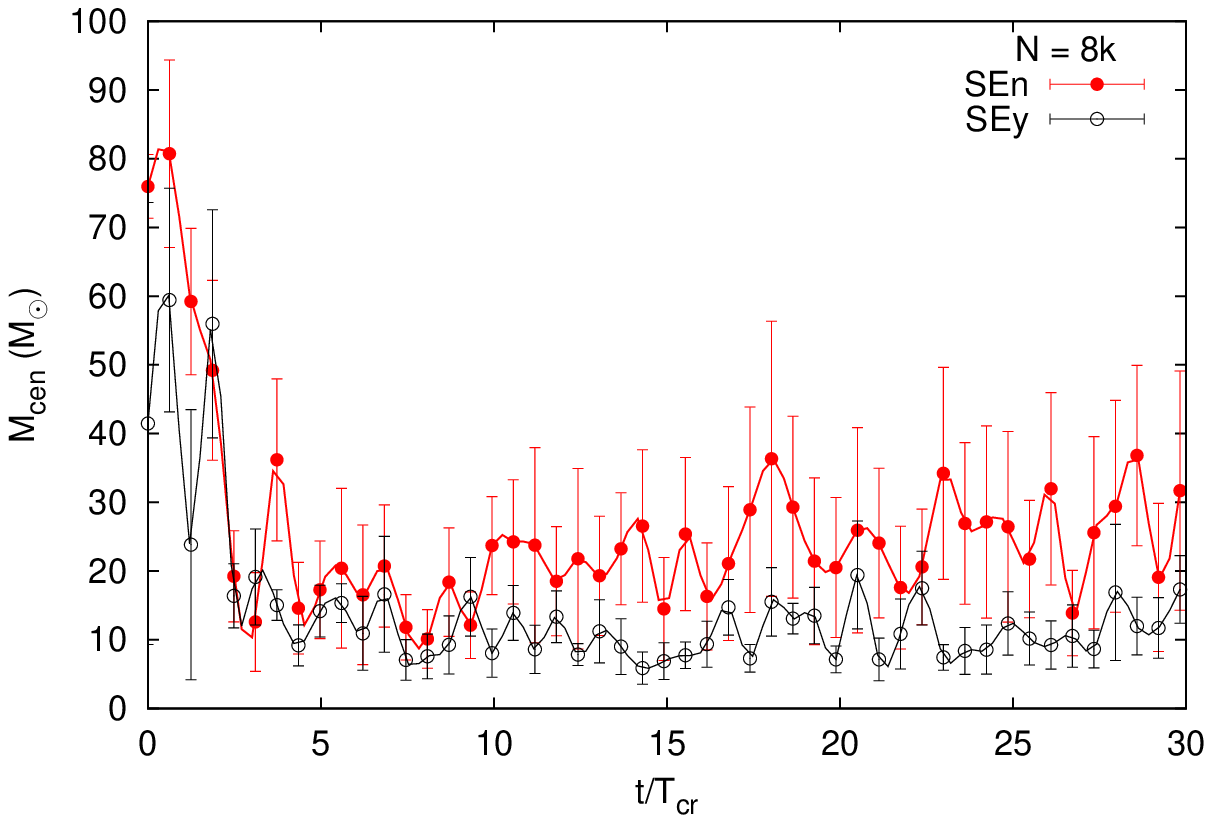}
\includegraphics[width = 8cm]{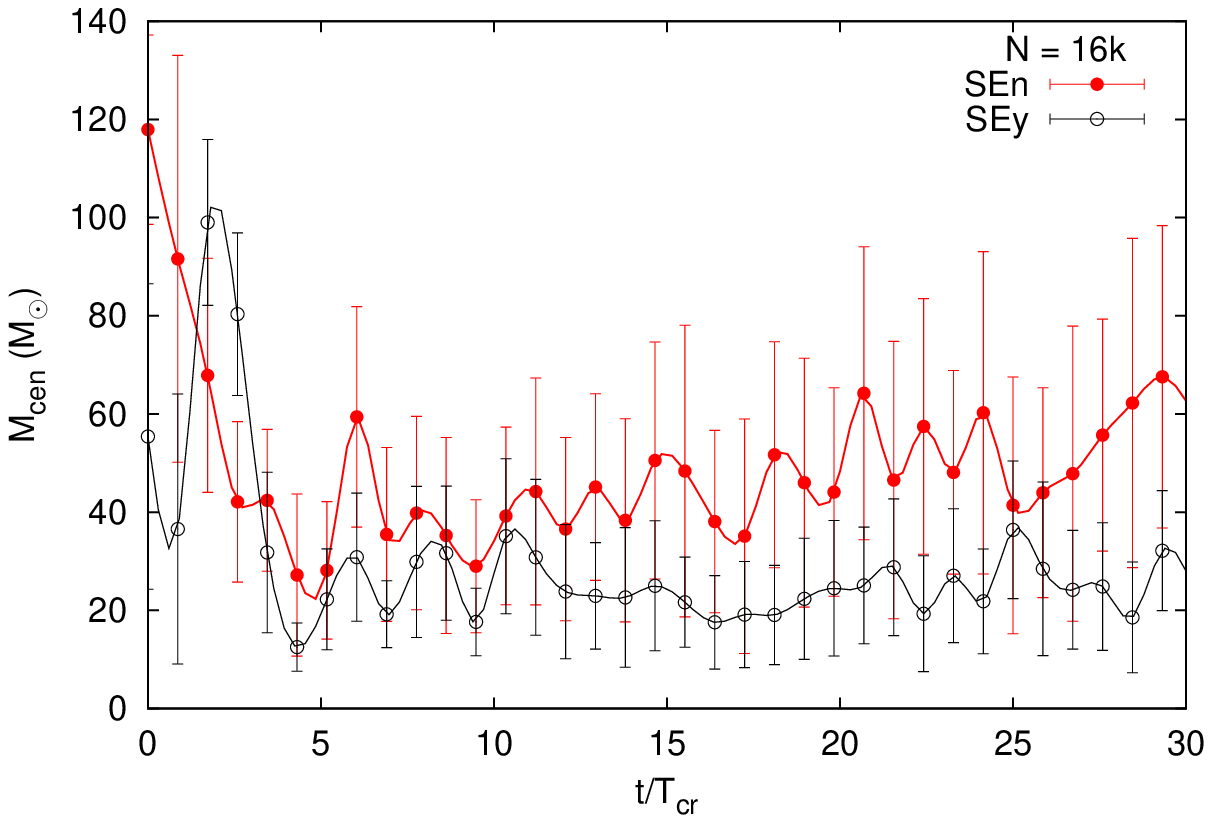}\\
\includegraphics[width = 8cm]{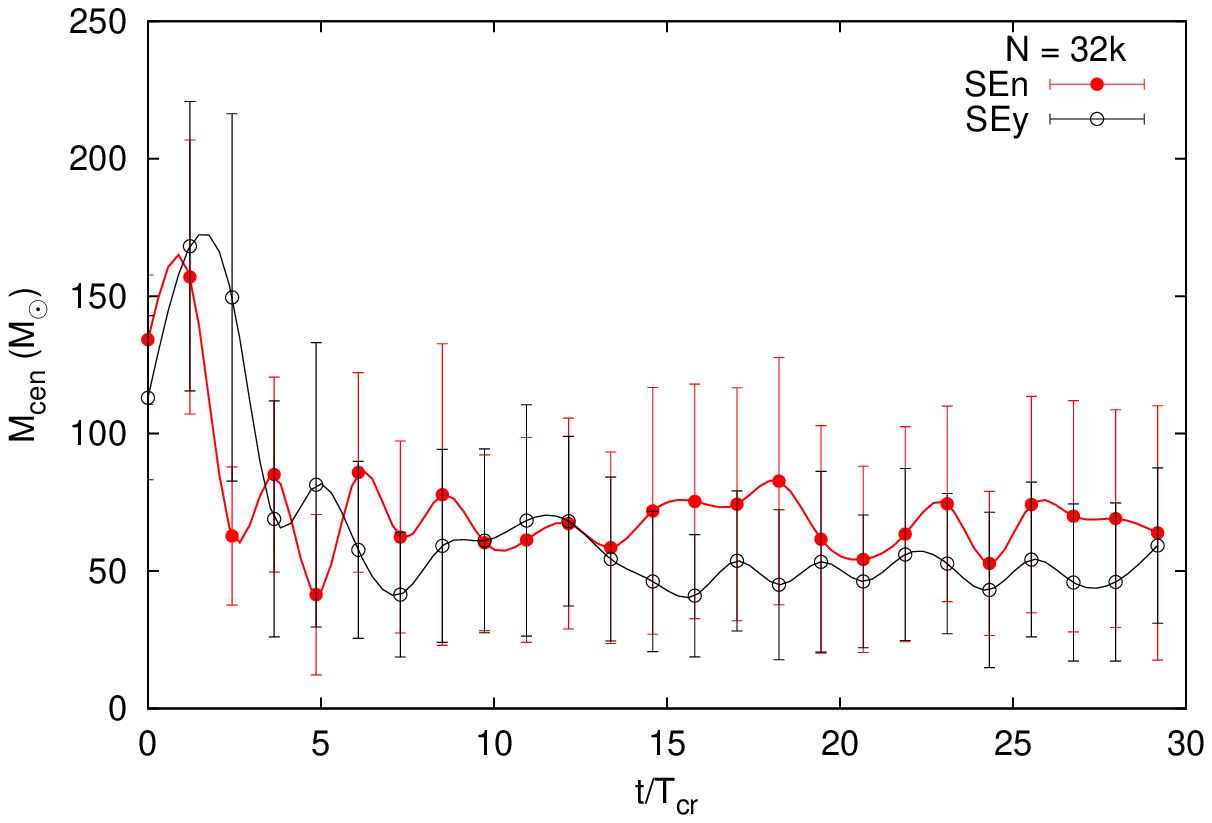}
\includegraphics[width = 8cm]{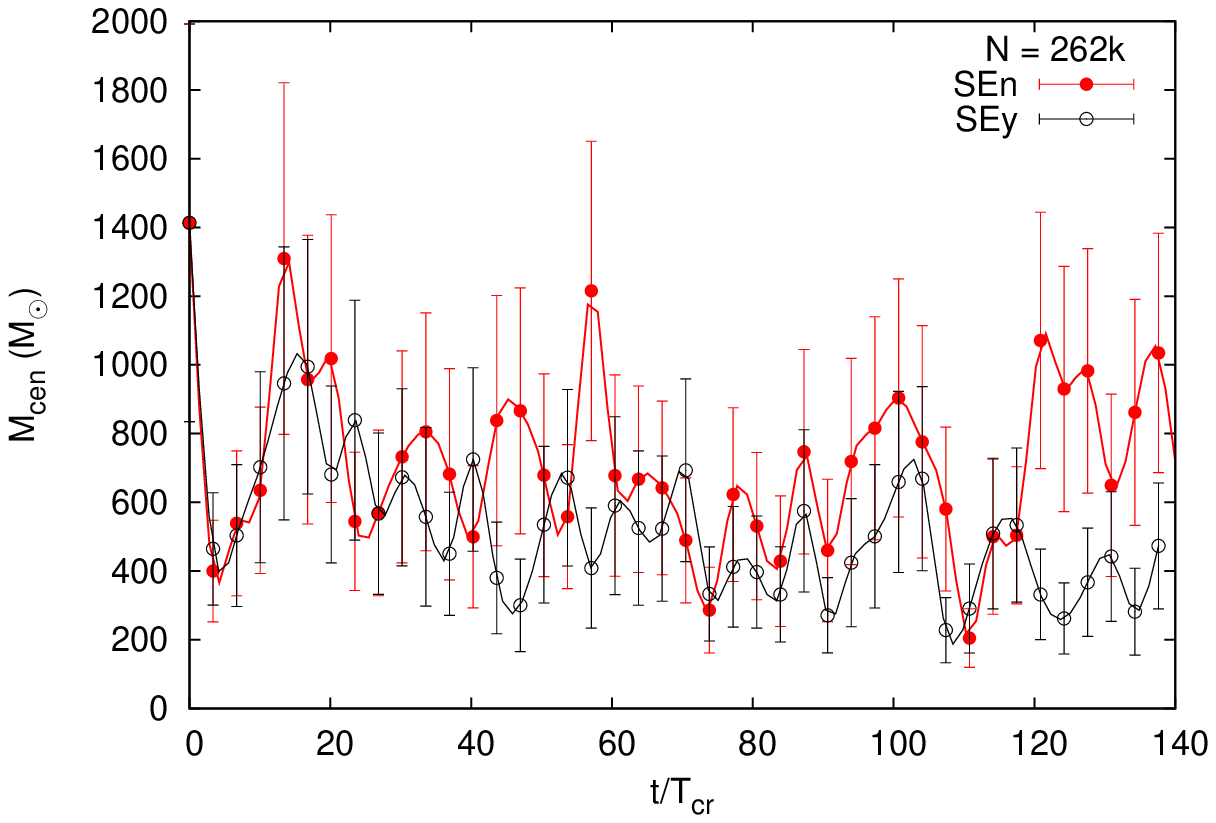}\\
\includegraphics[width = 8cm]{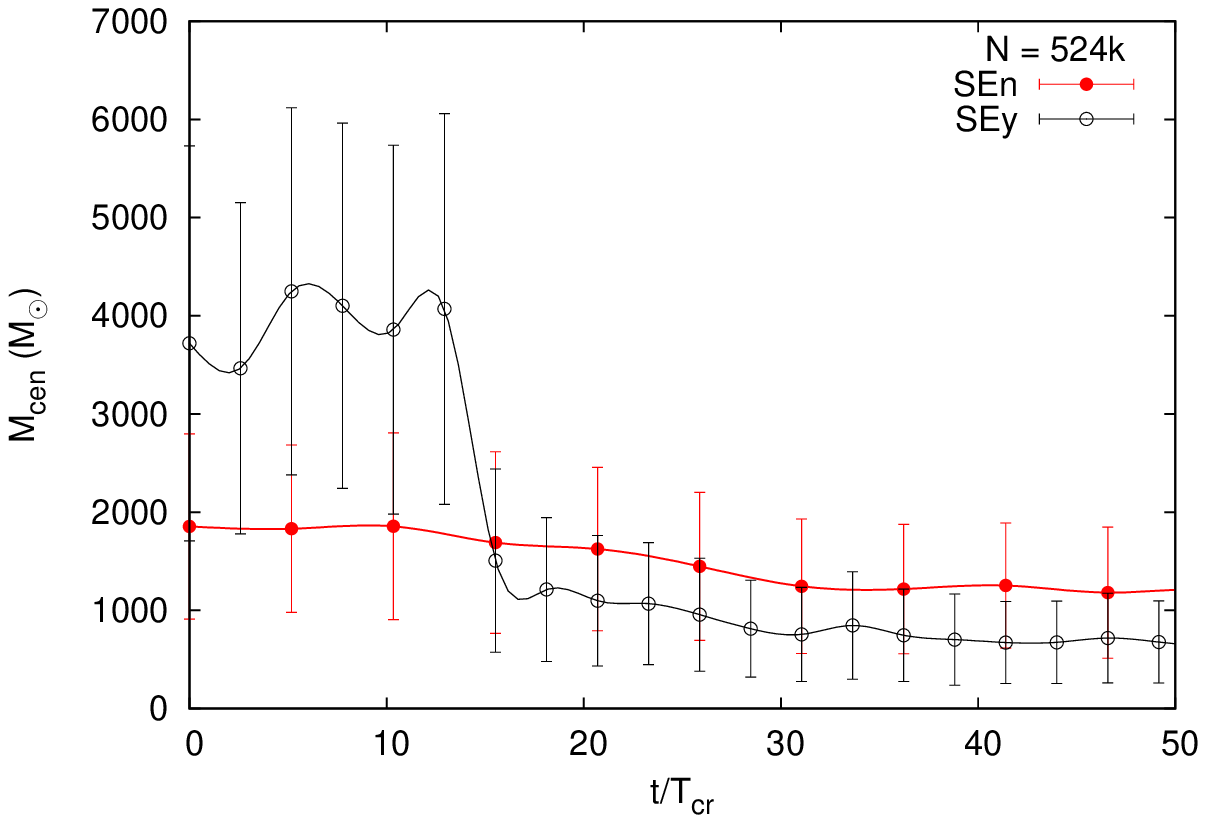}
\caption{Mass enclosed within $r_{\rm MSS}$ as a function of the time for all the simulations performed. Filled red circles represent simulations performed with HiGPUs, whereas open black circles represent simulations performed with HiGPUsSE.}
\label{mcent}
\end{figure*}

Figure \ref{mcent} shows the mass enclosed within $r_{\rm MSS}$ for all the simulations performed, $M_{\rm MSS}$. In particular, we compared in each panel simulations performed with HiGPUs (labelled with SEn), in which stellar evolution is evaluated ``a posteriori'' with respect to the dynamical evolution of the cluster, and those performed through HiGPUsSE (labelled with SEy), in which stellar evolution is accounted along with the dynamical evolution of the system.

Our estimate of the length scale of the host cluster is affected by a $\sim 20\%$ uncertainty, leading to a corresponding uncertainty on the value of $r_{\rm MSS}$ (see Equations \ref{r_st} and \ref{r_st2}).

As expected, the mass of MSSs in simulations labelled with SEn is slightly higher than 
in simulations denoted with SEy, since in the latter case mass lost by each star decreases the df efficiency, thus increasing its orbital decay time-scale. On the other hand, it is quite evident that the mass estimates provided using both HiGPUs and HiGPUsSE are comparable within the error. Hence, stellar evolution seems to not affect significantly the early phase of formation of MSSs.

Moreover, it should be noted that the scaling relations provided in Section \ref{Scor} allow to evaluate MSS masses that differ few percent from those evaluated through the $N$-body simulations. 

In the next section, we will focus the attention on the formation of an MSS in models composed of 262k and 524k particles simulated through HiGPUs, which allows a much faster integration with respect to HiGPUsSE.

\subsubsection{Formation of an MSS in globular clusters}

In configuration \texttt{262k\_SEn\_D\_S} the orbital decay of the heaviest stars leads to the formation of a dense sub-system, MSS, with a half-mass radius of $r_{\rm h} = 0.06$ pc, which extends up to $0.1$ pc. 

\begin{figure}
\centering
\includegraphics[width=8cm]{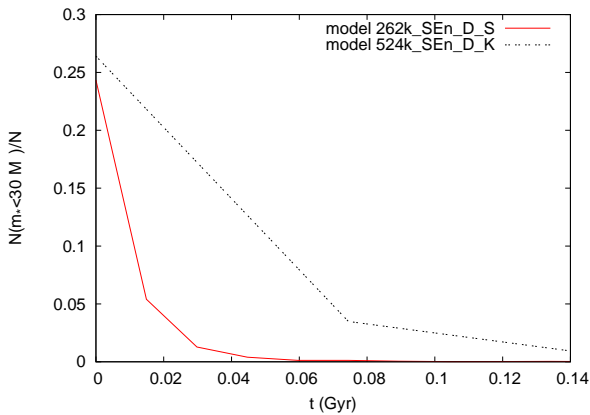}
\caption{Fraction of stars with masses below $30$ M$_\odot$ which move within the size of the MSS, as a function of the time, in models \texttt{262k\_SEn\_D\_S} (solid line) and  \texttt{524k\_SEn\_D\_K} (dotted line). In both the simulations the whole population of light stars is pulled out from the MSS.}
\label{light}
\end{figure}

As shown in Figure \ref{light}, the fraction of stars with initial masses below $30$ M$_\odot$ that move within $r_{\rm MSS}$ rapidly decreases in time both in simulations \texttt{262k\_SEn\_D\_S} and \texttt{524k\_SEn\_D\_K}, leading to the formation of an MSS mainly composed of heavy stars.

\begin{figure}
\centering
\includegraphics[width = 8cm]{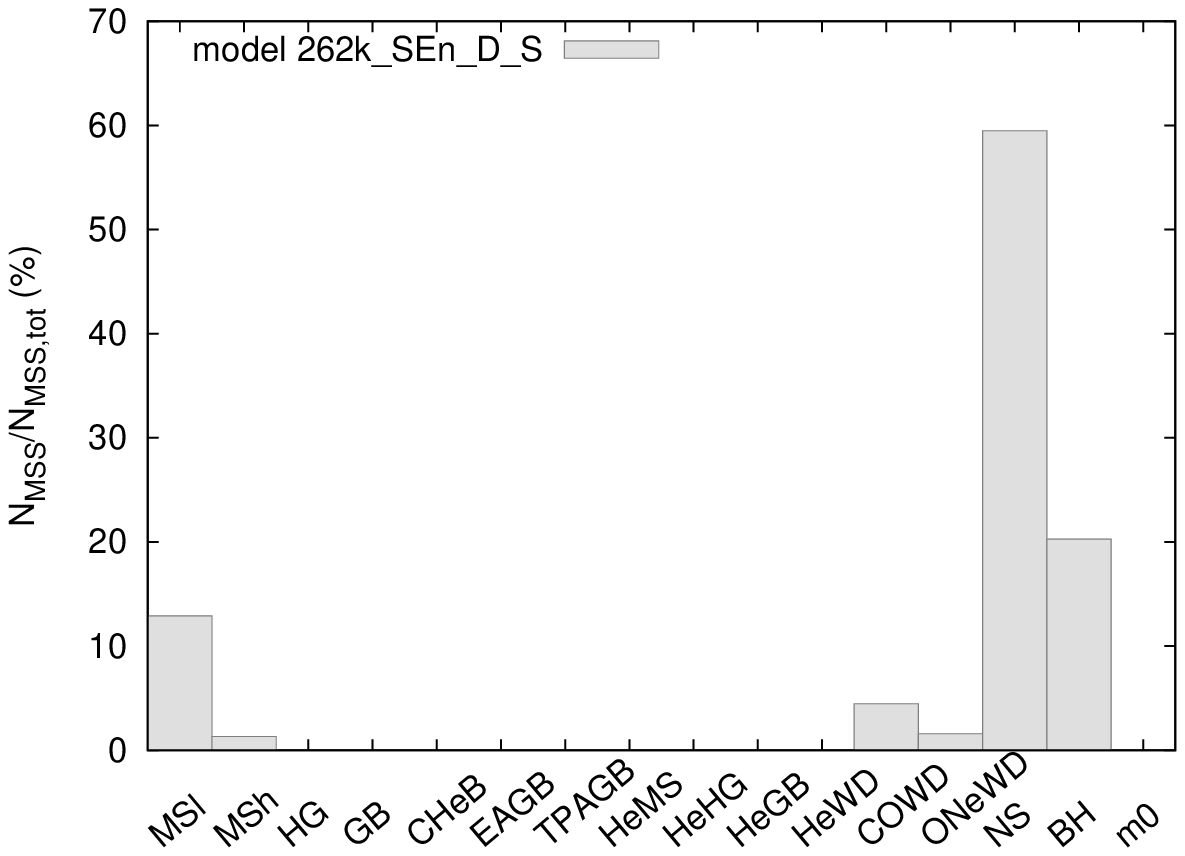}
\caption{As in Figure \ref{Fn4}, but for simulation \texttt{262k\_SEn\_D\_S} }
\label{Fn4NB}
\end{figure}

In particular, model \texttt{262k\_SEn\_D\_S} contains a population of $68$ stars with masses above $30$ M$_\odot$, for a total mass of $3163$ M$_\odot$, that drives the formation of an MSS over a time-scale of $20$ Myr.
Figure \ref{Fn4NB} shows the distribution of stellar types within the MSS in this case assuming an age of $2$ Gyr, in order to compare this results with the semi-analytical results discussed in the previous section. The MSS composition is really similar to the theoretical results shown in Figure \ref{Fn4}, with a dominant fraction of NSs and BHs and a small percentage of MS stars and WDs.

During the MSS assembly, heavy stars transfer energy to lighter particles, which move outward, leading the total energy of the MSS to decrease reaching negative values over a time $t\sim 65$ Myr. 
During this process, the MSS contracts continuously, the most massive stars form the innermost core of the MSS while lighter stars move outward. Due to this, the long-term evolution of the MSS depend only on the gravitational encounters among the most massive stars, which may halt the MSS contraction through the formation of binary systems and close encounters.

On the other hand, it has been widely shown that close encounters and binaries should not affect significantly the global properties of the MSS \citep{heggie14,fregeau06}. In particular, \cite{morscher15} showed that strong interactions among binaries and single stars should not alter significantly the fraction of stellar BHs.

The energy provided by the binary hardening leads the cluster to expand by a factor of $\sim 2-2.5$, depending on the metallicity of the host \citep{trani14}. 
Using Equations \ref{r_st} and \ref{r_st2}, and accounting for this expansion the values of $r_{\rm MSS}$ presented above still agrees with observational estimates  \citep{vandermarel10,haggard13,lanzoni13}.

Assuming for this model a value of the metallicity $Z=0.0004$, we found from the $N$-body simulation that the newly born MSS should have a mass $M_{\rm MSS}=500 \pm 80$ M$_\odot$. On the other hand, making use of the scaling relations provided above for a similar model, we found $M_{\rm MSS}=513 \pm 15$ M$_\odot$, a value compatible with the one obtained through the $N$-body simulation.

The formation process of an MSS can be argued also looking at the evolution of the host cluster's structure. For instance, we found that its surface density profile $\Sigma(R)\propto R^{-\alpha}$, with $R\lesssim 0.1$ pc the projected distance to the cluster centre, 
gets steeper as the MSS formation proceeds, and the slope passes from $\alpha=1$ to $\alpha\simeq1.30\pm 0.05$ by the end of the simulation. Moreover, we found an increasing slope in the projected velocity dispersion, $\sigma(R)\propto R^{-\beta}$. Indeed, we found that $\beta$ passes from a nearly $0$ value to $0.12\pm0.01$ within $R\lesssim 0.1$ pc. This findings agree with observations, since the observed mass excesses are often associated to power-law surface density profiles or projected velocity dispersions within the inner region of the host clusters \citep{vandermarel10,noyola10,Kamann14}.

Regarding configuration \texttt{524k\_SEn\_D\_K}, instead, the population of stars heavier than $30$ M$_\odot$ has an initial mass of $M=2.7\times 10^4$ M$_\odot$ and is composed of $561$ particles.

In this case, mass segregation process is slower than in the previous model, as expected by Equation \ref{eq2} (see Figure \ref{lagr}). Furthermore, it should be noted that the stellar mass loss time-scale in model \texttt{524k\_SEn\_D\_K} is smaller than the segregation time-scale. Due to this, stars will reach the GC centre after they have lost most of their initial mass and, therefore, the subsequent evolution of the MSS will be substantially dominated by binary formation and close encounters.

The concentration of the stars leads to an MSS with a half-mass radius $0.1$ pc, slightly larger than in the case \texttt{262k\_SEn\_D\_S}. It is worth noting that, even in this case, the MSS size estimates provided in Section \ref{proMSS} are comparable to the size achieved through the $N$-body representation. 

As in model \texttt{262k\_SEn\_D\_S}, the fraction of stars with initial masses below $30$ M$_\odot$ approaches a value close to $0$ within the simulated time, leaving an MSS composed only of the most massive stars. 

For this model, assuming a metallicity $Z=0.0004$, we found that the MSS should have a mass $M_{\rm MSS}\simeq 1960$ M$_\odot$, a value $7\%$ smaller than the value estimated through the scaling relations provided above.

\section{Discussion and conclusions}
\label{end}

In this paper we investigated how the global properties of star clusters can affect the formation process, in their centres, of massive sub-systems (MSSs), using a semi-analytical method and a series of direct $N$-body simulations.

Despite the methodology used here do not account for the effects related to the formation of binary and multiple systems, which may alter significantly the dynamics of stars in the cluster, it should noted that only hard binaries composed of compact objects (stellar BHs or NSs) may inject enough energy to halt the contraction \citep{chernoff96,heggie75,fregeau06}. Moreover, recent works have shown that the formation of binaries should not deplete the population of compact stars within the cluster centre \citep{mackey08,heggie14,morscher15}. Hence, the formation of binaries within the cluster centre should not affect significantly our estimates of MSS masses and size, which represent reliable upper limits to the observed mass excesses, despite the simplicity of our approach.

In the following, we summarise our main results:
\begin{itemize}
\item in order to understand how the global properties of a star cluster affect the formation of an MSS, we investigated mass segregation in 168 models of star clusters with different initial density distribution, total mass, IMF and metallicity, aiming to quantify their role;
\item we found that the mass growth of an MSS is characterised by three distinct phases: i) a fast phase during which $M_{\rm MSS}$ increases through orbital decay of heavy stars with small apocentres; ii) a second phase during which $M_{\rm MSS}$ decreases as a consequence of stellar mass loss; iii) a slow phase during which $M_{\rm MSS}$ smoothly rises in corrispondence of the occasional orbital decay of stars that move in a peripheral region of the cluster;
\item though our results indicate that nor the spatial distribution neither the metallicity play a significant role in determining the final mass of an MSS, they are much more effective in determining which kind of stars populate it. Indeed, we found that MSSs formed in GCs characterised by an initial uniform density distribution should host a significant fraction of MS stars, while in the case of Dehnen density profiles the MSSs are populated mainly by NSs and BHs. Furthermore, we found that low metal MSSs contains more BHs than NSs, unlike MSSs with solar metallicity;
\item we have shown that the population of stellar BHs tends to form a tight nucleus, which extends up to $0.1$ pc, quite independently on the GC properties. On the other hand, the distribution of other stellar type is more influenced by the spatial distribution of the host;
\item we provided scaling relations connecting the masses of the MSSs and their host clusters, showing that they are in overall agreement with previous theoretical and observational works. In particular, we have shown that these correlations allow to obtain a value for the central mass excess of clusters $\omega$ Cen in excellent agreement with the observed values provided by \cite{vandermarel10};
\item in order to test our semi-analytical approach we performed a series of direct $N$-body simulations using the code HiGPUs. Moreover, we used HiGPUsSE, a modified version of HiGPUs in which we included stellar evolution, in order to highlight its effects on the MSSs formation process;
\item comparing these simulations with our theoretical approach, we demonstrated that the treatment for df used here is well suited to describe the orbital decay of massive stars. Moreover, comparing the results obtained with HiGPUs and those obtained through HiGPUsSE, we found that stellar evolution seems to not alter significantly the dynamical evolution of the cluster nucleus, where the MSS form;
\item we have shown that our treatment allows to predict quite precisely which kind of stars will populate the MSS, as shown through the comparison with the $N$-body runs;
\item the global properties of the host cluster reflect the ongoing formation of an MSS. For instance, we observed in \texttt{262k\_SEn\_D\_S} a significant increase in the slope of the surface density profile of the cluster and in the slope of the velocity dispersion;
\item the agreement found between our semi-analytical estimates and the direct $N$-body simulations indicates that the procedure followed in this paper can be used to provide reliable initial conditions for simulating the long-term evolution of GCs. This would allow to significantly shorten the computational time needed to model the long-term evolution of a GC, since the completion of the mass segregation process in a typical GC would require computational times that exceed one month.
\end{itemize}

\section*{Aknowledgement}

I warmly thank R. Capuzzo-Dolcetta for discussions and comments about the contents investigated in this paper, M. Donnari for the support provided during the paper writing and M. Spera for his unvaluable support provided to merge the SSE package in the HiGPUs code.
I am also grateful to B. Lanzoni, A. Gualandris, D. Heggie, S. Portegies-Zwart, M. Mapelli, N. Lutzgendorf and to the anonymous referee for their useful suggestions that helped me to improve an earlier version of this manuscript. 
The author aknowledge the Italian Ministry of Education, University and Research (MIUR), which funded part of this work through the grant PRIN 2010 LY5N2T 005. The research presented here has been developed in the framework of the project ``the MEGAN project: Modelling the Evolution of GAlactic Nuclei'', funded from the University of Rome ``Sapienza'' through the grant D.D. 52/2015.

\footnotesize{
\bibliographystyle{mn2e}
\bibliography{bblgrphy}
}

\end{document}